\documentclass[12pt,preprint]{aastex62}
\usepackage{natbib}
\usepackage{xcolor}

\newcommand{\eps}[1]{\mbox{log~$\epsilon$(#1)}} 
\newcommand\species[2]{#1~{\sc #2}}

\def\eg{\mbox{e.g.}}

\def\rpro{\mbox{$r$-process}}

\def\ncap{\mbox{$n$-capture}}
\def\teff{\mbox{T$_{\rm eff}$}}
\def\logg{\mbox{log~{\it g}}}
\def\vmicro{\mbox{$\xi_{\rm t}$}}
\def\kmsec{\mbox{km~s$^{\rm -1}$}}
\def\hdeight{\mbox{HD 84937}}
\def\bdzero{\mbox{BD$+$03$^{\rm o}$740}}
\def\bdone{\mbox{BD$-$13$^{\rm o}$3442}}
\def\cdthree{\mbox{CD$-$33$^{\rm o}$1173}}

\shorttitle{Stellar Fe-Group Abundances}
\shortauthors{Cowan et al.}

\begin{document}

\title{Detailed Iron-Peak Element Abundances in Three Very  Metal-Poor
       Stars\footnote{%
Based on observations made with the NASA/ESA
\textit{Hubble Space Telescope},
obtained at the Space Telescope Science Institute (STScI), which is
operated by the Association of Universities for
Research in Astronomy, Inc.\ (AURA) under NASA contract NAS~5-26555.
These observations are associated with program GO-14232.
Some
data presented in this paper were obtained from the
Barbara A.\ Mikulski Archive for Space Telescopes (MAST).~
These data are associated with Program GO-7402.
Other data have been obtained from the European Southern Observatory (ESO)
Science Archive Facility.
These data are associated with Programs
67.D-0439(A),
68.B-0475(A),
68.D-0094(A), and
095.D-0504(A).
This research has also made use of the Keck Observatory Archive (KOA),
which is operated by the W.M.\ Keck Observatory and
the NASA Exoplanet Science Institute (NExScI),
under contract with NASA.
This work has also made use of data collected from
McDonald Observatory of the University of Texas at Austin.
 }}

\author{John J. Cowan}
\affiliation{Homer L. Dodge Department of Physics and Astronomy, University 
             of Oklahoma, Norman, OK 73019; jjcowan1@ou.edu}
\author{Christopher Sneden}
\affiliation{Department of Astronomy and McDonald Observatory, The University 
             of Texas, Austin, TX 78712; chris@verdi.as.utexas.edu}
\author{Ian U. Roederer}
\affiliation{Department of Astronomy, University of Michigan, 
             1085 S. University Ave., Ann Arbor, MI 48109, USA; iur@umich.edu}
\affiliation{Joint Institute for Nuclear Astrophysics -- Center for the
Evolution of the Elements (JINA-CEE), USA}
\author{James E. Lawler}
\affiliation{Department of Physics, University of Wisconsin-Madison,
             1150 University Ave., Madison, WI 53706; jelawler@wisc.edu,
             eadenhar@wisc.edu}
\author{Elizabeth A. Den Hartog}
\affiliation{Department of Physics, University of Wisconsin-Madison,
             1150 University Ave., Madison, WI 53706; jelawler@wisc.edu,
             eadenhar@wisc.edu}
\author{Jennifer S. Sobeck}
\affiliation{Department of Astronomy, University of Washington, Seattle, 
             WA 98195, USA;  jsobeck@uw.edu}
\author{Ann Merchant  Boesgaard}
\affiliation{Institute for Astronomy, University of Hawai’i at Manoa, 
             2680 Woodlawn Drive, Honolulu, HI 96822, USA; boes@ifa.hawaii.edu}

\begin{abstract}
We have obtained new detailed abundances of the Fe-group elements Sc through 
Zn (Z~=~21$-$30) in three very metal-poor ([Fe/H] $\approx -3$) stars: 
\bdzero, \bdone,  and \cdthree.
High-resolution ultraviolet 
\textit{HST}/STIS spectra in the wavelength range 
2300$-$3050~\AA\ were gathered, and complemented by an assortment of optical 
echelle spectra.
The analysis featured  recent laboratory atomic data for number 
of neutral and 
ionized 
species for all Fe-group elements except Cu and Zn.
A detailed examination of scandium, titanium, and vanadium abundances
in large-sample spectroscopic surveys indicates that they are positively 
correlated in stars with [Fe/H]~$<$~$-$2. 
The abundances of these elements 
in \bdzero, \bdone,  \cdthree, and \hdeight\ (studied in a previous paper
of this series) are in accord with these trends and lie at the high end of 
the correlations. 
Six elements 
have
detectable neutral and ionized features, and generally their 
abundances are in reasonable agreement.  
For Cr we find only minimal abundance disagreement between the neutral 
(mean of [Cr~\textsc{i}/Fe]~$=+0.01$) and ionized
species (mean of [Cr~\textsc{ii}/Fe]~$=+0.08$), unlike most studies in the past.
The prominent exception is Co, for which the neutral species indicates
a significant overabundance 
(mean of [Co~\textsc{i}/H]~$=-2.53$), 
while no such enhancement is seen for the ionized
species (mean of [Co~\textsc{ii}/H]~$=-2.93$).
These new stellar abundances, especially the correlations among Sc, Ti, 
and V, suggest that models of element production in early high-mass metal-poor
stars should be revisited.

\end{abstract}


\section{INTRODUCTION\label{intro}}

The origin of the elements is one of the important questions in modern 
astrophysics.
With the exception of a few elements formed in the big bang, most all 
elements are synthesized in stars. 
But astrophysical sites and nucleosynthetic conditions
that produce individual, as well as groups of, elements vary widely.
Of particular interest to nucleosynthesis studies are relatively
metal-poor stars of the Galactic halo, whose abundance sets must have
been generated by few prior massive star deaths.

In the early 2000's our group concentrated on determining accurate abundances 
of neutron-capture elements ($Z > 30$, hereafter \ncap), particularly the 
rare-earth (RE) elements in metal-poor halo stars, in an attempt to 
understand the origin 
of these elements (\citealt{cowan19} and references therein).  
Our approach relied upon two major efforts. 
First, we obtained high resolution spectra, both in the visible and 
ultraviolet, utilizing both ground-based and space-based telescopes, to 
identify as many \ncap\ transitions as possible.
Second, we used Wisconsin 
laboratory atomic physics 
studies to obtain precise transition probabilities and hyperfine/isotopic 
substructures to greatly increase the reliability of individual species 
abundances.
This multi-year effort to produce accurate stellar \ncap\ elemental 
abundances culminated in \cite{lawler09} and \cite{sneden09}, and references 
therein. 

More recently we have turned attention to the iron-group elements 
($Z$~=~21$-$30), because their abundances in metal-poor stars can be directly
compared to predicted outputs from massive star element donors.
Such studies face difficulties. 
First, for many stars not all Fe-group elements are easily detectable with
the available spectra of low metallicity stars.
Secondly, even when elements can be detected, often only neutral-species 
transitions are available for analysis in typical ground-based spectra.
The populations  of Fe-group elements in warmer dwarfs and 
cooler red giants usually are dominated by the ionized species;
therefore the elemental abundances are mostly based on results from
the minority species.
Thirdly, in many instances only a handful of lines are employed to derive 
individual elemental abundances.
Finally, significant questions have been raised about whether local 
thermodynamic equilibrium (LTE) can adequately describe the ionization 
equilibrium, and large upward corrections have been proposed to LTE-based 
abundances from neutral species (e.g., \cite{bergemann10}, \cite {andrievsky18b} and \cite{shi18}).

Our group has been concentrating on improving the laboratory data for
Fe-group neutral and singly-ionized transitions normalized with radiative
lifetimes from laser induced fluorescence.  Lines of interest  to cool-star
stellar spectroscopy are emphasized.
Citations to individual papers within this series will be given in 
\S\ref{newabunds}.
In each study we have used the new atomic data to re-derive abundances
in the solar photosphere and in the very metal-poor 
([Fe/H]~$\sim$~$-$2.2\footnote{
We adopt the standard spectroscopic notation \citep{wallerstein59} that for 
elements A and B,
[A/B] $\equiv$ log$_{\rm 10}$(N$_{\rm A}$/N$_{\rm B}$)$_{\star}$ $-$
log$_{\rm 10}$(N$_{\rm A}$/N$_{\rm B}$)$_{\odot}$.
We use the definition 
\eps{A} $\equiv$ log$_{\rm 10}$(N$_{\rm A}$/N$_{\rm H}$) + 12.0, and
equate metallicity with the stellar [Fe/H] value.
Also, \eps{\species{X}{i}} or \eps{\species{X}{ii}} are to be understood as  
an elemental abundance determined from the named species.})
main sequence turnoff star \hdeight.
This star has been chosen because of its brightness, well-known atmospheric
parameters, and availability of good high-resolution spectra over a very
large wavelength range (see \citealt{sneden16}, hereafter SN16).  

In this paper we expand our analyses of the Fe-group elements to include 
three very metal-poor ([Fe/H]~$\simeq$~$-$3) stars, \bdzero,  
\bdone, and \cdthree.
That study reported many have not been 
This represents the first-ever study of the Fe group with recent, 
consistent neutral and ionized species lab data for elements Sc through Ni 
in such low-metallicity halo stars, which should provide probes of early 
Galactic nucleosynthesis.

\textbf{Our three warm, main-sequence turnoff stars are well-known targets 
of low-metallicity chemical composition studies. 
Many high-resolution spectroscopic investigations have centered on questions 
about their Li abundances and isotopic ratios (\eg, \citealt{ryan99}, 
\citealt{bonifacio07}, \citealt{hosford09}, and references therein).
There have been few recent comprehensive abundance analyses, caused in large 
part by the extreme line-weakness of their optical spectra.
A significant exception is the detailed study of 28 main-sequence, subgiant, 
and giant stars in the metallicity range $-$3.7~$\leq$~[Fe/H]~$\leq$~$-$2.5
by \cite{lai08} with spectra that extended into the near-UV spectral
domain ($\lambda$~$\gtrsim$~3100~\AA).
That work included \bdzero, for which \citeauthor{lai08} reported abundances
for 15 species of the 10 Fe-group elements.
We will compare our results with those of \citeauthor{lai08} in 
\S\ref{newabunds}.}

\cite{roederer18}, hereafter Paper~1, derived model atmospheric parameters
and metallicities from $\gtrsim$250 \species{Fe}{i} and \species{Fe}{ii} 
transitions in these stars.
Here we present abundances for the other Fe-group elements.
In \S\ref{specdata} we introduce the high-resolution spectroscopic
data sets analyzed for the three stars.
New abundance determinations of several elements and summaries of recent
results for other elements are given in \S\ref{newabunds} with final 
iron-peak elemental abundances listed in \S\ref{finalabs}.
Interpretation and discussion of the iron-peak data and abundances,  
particularly nucleosynthesis origins and production mechanisms, 
along with the implications for Galactic chemical evolution,  
are included in \S\ref{results}.  
Finally,  a summary and conclusions are detailed in \S\ref{sum}.

\section{SPECTROSCOPIC DATA AND ANALYSES\label{specdata}}

The high resolution spectra for \bdzero,  \bdone, and \cdthree \ (hereafter
called ``program stars'') have been described in detail in Paper~1. 
Briefly, the spectra in the UV region (2290$-$3050~\AA) were
gathered with the 
Space Telescope Imaging Spectrograph (STIS; 
\citealt{kimble98,woodgate98}) 
on board the \textit{Hubble Space Telescope} (\textit{HST}).~
The instrumental configuration was with the E230M echelle grating centered 
at $\lambda$2707, the 0\farcs06~$\times$~0\farcs2 slit, and
the NUV Multianode Microchannel Array (MAMA) detector.
The spectral resolving power was $R$~$\equiv$~$\lambda/\Delta\lambda$
$\simeq$~30000.
The \textit{HST}/STIS spectra for the three program stars
were obtained with STIS in Program GO-14232. 
See Paper~1 for discussion of the observations of \hdeight.

Optical spectra of these stars were obtained from online archives from
the ESO Very Large Telescope Ultraviolet and Visual Echelle Spectrograph 
(UVES; \citealt{dekker00}), the Keck~I Telescope High Resolution Echelle 
Spectrometer (HIRES; \citealt{vogt94}), and the McDonald Observatory 
Harlan J.\ Smith Telescope and Robert G.\ Tull Coud\'{e} 
Spectrograph \citep{tull95}.
All stars have optical spectra with $R$~$\gtrsim$~40,000 and wavelength
coverage mostly complete in the 3050$-$6800~\AA\ range.
See Table~1 of Paper~1 for the characteristics of all the spectra used in
our work.

\section{ABUNDANCE DETERMINATIONS FOR \bdzero,  \bdone, 
AND \cdthree\label{newabunds}}

The analyses followed the methods used for our study of \hdeight\ 
(SN16).
Derivation of model atmospheric parameters was done with Paper~1 and
will not be repeated here.  
We adopted the values from that paper for the program stars.
For \hdeight\ we used the model parameters from SN16.
The parameters for these four stars are listed in Table~\ref{tab-models}.
Stellar 
photospheric 
models with these parameters were interpolated
from the \cite{kurucz11} model grid\footnote{
http://kurucz.harvard.edu/grids.html} using software kindly provided by
Andrew McWilliam and Inese Ivans.

In this study all line abundances were determined with interactive fits
of observed and synthetic spectra.
The linelists for generating the synthetic spectra were as described
in SN16.
Beginning with the \cite{kurucz11} atomic and molecular line 
database\footnote{
http://kurucz.harvard.edu/linelists.html}, we added/substituted/corrected the
atomic spectral features that have been included in recent accurate laboratory 
data papers by the Wisconsin atomic physics and Old Dominion molecular physics
programs (\eg, \citealt{yousefi18}, \citealt{lawler19}, and previous 
studies in those series).\footnote{
A code to accomplish this is available at https://github.com/vmplacco/linemake,
along with all the transitions used by that code.  A list of the laboratory
studies contributing to this database can be found at 
http://www.as.utexas.edu/chris/lab.html .}
These linelists and the stellar atmospheric models were used by the LTE
plane-parallel synthetic spectrum code MOOG \citep{sneden73}. 
We used the code version that includes scattering in the continuum source
function \citep{sobeck11}, although in our warm main-sequence turnoff stars
the scattering contribution is very small even in the UV spectral
region.

\begin{figure}
\epsscale{1.0}
\plotone{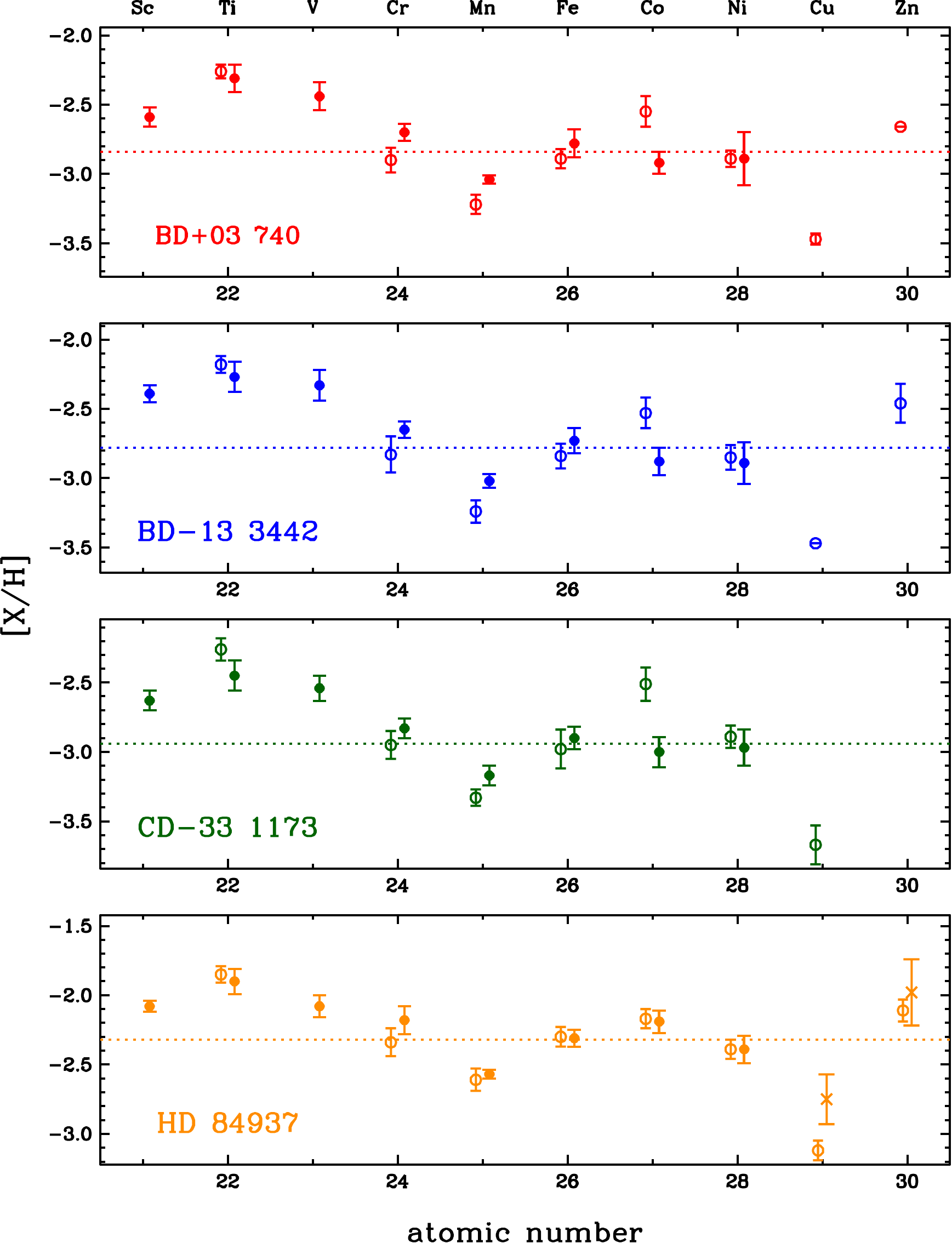}
\caption{
\label{fig1} \footnotesize
Abundance ratios [X/H] for the three program stars and for \hdeight.
In each panel the dotted line represents the mean [Fe/H] for that star.
The abundances derived from neutral-species transitions are shown as open
circles, and they are shifted slightly to the right of their atomic numbers.
The abundances from ionized-species transitions are shown as filled circles 
and shifted slightly to the left. 
In the panel for \hdeight, we show the Cu~\textsc{ii} 
and Zn~\textsc{ii} abundances from 
\cite{roederer18a} as $\times$ symbols.
\textbf{The error bars here are the sample standard deviation $\sigma$ 
values from Table~\ref{tab-abmeans} for each species abundance.}
}
\end{figure}

In Table~\ref{tab-lines} we list all the transitions used in our analyses,
and the resulting abundances for the three program stars.
Mean abundances in both log~$\epsilon$ and [X/H] forms
are given in Table~\ref{tab-abmeans}, along with their line-to-line standard 
deviations. 
\textbf{However, \species{Cu}{i} and \species{Zn}{i} are represented by only 
two transitions each per star, and in several cases the species statistical 
$\sigma$ values are unrealistically near 0.00.
Therefore in such cases we have entered ``(0.1)'' for their standard 
deviations in Table~\ref{tab-abmeans}.}
The mean abundances and standard deviations are illustrated in 
Figure~\ref{fig1}, along with the values for \hdeight\ (taken from 
SN16).

As a general rule, we were able to detect the most lines for \bdzero,
the next most for \bdone, and the fewest for \cdthree, approximately in
order of increasing \teff\ (which produces weaker-lined spectra for a given
metallicity) and slightly decreasing signal-to-noise ($S/N$) ratios in the 
\textit{HST}/STIS spectra
of these stars.
The abundances for \hdeight\ in this table are quoted from SN16.

In our stars the Fe-group elements Sc through Ni ($Z$~=~21$-$28) have many
detectable ionized-species transitions, all of which have benefited from
recent lab studies.
Additionally, in the atmospheres of our stars these elements are almost
completely ionized, as shown in Figure~2 of SN16.
The neutral species account for at most a few percent of the elemental
abundances.
Therefore our primary Sc$-$Ni abundance sources will be the ionized species,
while the neutrals will mainly serve as indicators of how well the basic
Saha balances hold with our LTE analytical approach.

\textbf{The only previous investigation of our stars to extensively examine 
their ionized-species transitions was that of \cite{lai08}, who reported 
abundances for Fe-group ionized species of Sc, Ti, V, Cr, Mn, and Fe in
\bdzero.
Our results to be discussed below  generally are in good accord with theirs,
with $\langle$[X/Fe]$_{Lai08}$~$-$ [X/Fe]$_{this work}\rangle$~=~0.02
for 14 neutral and ionized species in common.
We comment specifically on Cr in \S\ref{crab}.}

\subsection{Comments on Individual Species}\label{abspecies}

Here we will discuss some analytical aspects of each of the Fe-group elements,
generally deferring abundance interpretations to later sections.

\subsubsection{Scandium}\label{scab}
A new laboratory transition probability and hyperfine structure study for
\species{Sc}{i} and \species{Sc}{ii} has been published by \cite{lawler19}.
No \species{Sc}{i} lines in \hdeight\ were detectable by SN16.
Our program stars have similar temperatures and gravities to
those of \hdeight\ but $\simeq$~0.7~dex lower metallicities
(Table~\ref{tab-models}).
Thus it is not surprising that we failed to detect \species{Sc}{i} in
these stars also.
The series of Fe-group transition data papers has used \hdeight\ as
a test object for application to stellar spectra.
\cite{lawler19} derived log~$\epsilon$~= 1.08 ($\sigma$~=~0.05) from 29     
\species{Sc}{ii} lines.                                       
This new value is identical to that given for \hdeight\ in    
Table~\ref{tab-abmeans} from SN16.

\subsubsection{Titanium}\label{tiab}
Transition probabilities for \species{Ti}{i} are from \cite{lawler13} and   
for \species{Ti}{ii} are from \cite{wood13}.                  
These were also employed by SN16 for \hdeight.                
In the \citeauthor{wood13}\ paper attention was brought to a dip in abundances
from \species{Ti}{ii} in the blue  spectral region.            
For $\lambda$~$>$~3800~\AA\ the \hdeight\ mean abundance was  
log~$\epsilon$~= 3.12 ($\sigma$~=~0.04), in agreement with the mean value   
from \species{Ti}{i} lines.                                   
But for shorter wavelengths the \species{Ti}{ii} lines yielded
log~$\epsilon$~= 3.06 ($\sigma$~=~0.09), with the lowest values coming from 
lines with branching fractions greater than 0.2 in the wavelength region
most affected by the Balmer continuum opacity ($\sim$3100$-$3650~\AA);      
see \S7 of \citeauthor{wood13}, where this issue is discussed in detail     
without a clear resolution to the problem.                    
This issue 
does not appear to be significant
in our program stars, with a 
mean of the differences of the abundances 
from lines in the 3100$-$3800~\AA\ range minus those from lines 
with wavelengths $>$3800~\AA\ 
being $-$0.07~dex (Table~\ref{tab-abmeans}) compared to 
$-$0.06~dex in \hdeight.        
Our abundances from the ionized-species lines are $\simeq$~0.10~dex smaller  
than those from neutral-species ones, with \cdthree\ having the largest
species mismatch.

\subsubsection{Vanadium}\label{vab}
Using \species{V}{ii} transition probabilities and hyperfine structure      
data from \cite{wood14a} we determined V abundances from 30--47 lines      
in the program stars.                                         
Their mean abundance is log~$\epsilon$~$\simeq$~1.5, about 0.4~dex smaller  
than found for \hdeight\ (Table~\ref{tab-abmeans}).           
Unfortunately we could not check the ionization equilibrium 
for this element
because we were unable to detect \species{V}{i} 
lines in any of our program stars.
The neutral species has well-determined lab data for more than 800 lines    
(\citealt{lawler14} and \citealt{wood18}), but application to \hdeight\     
by \citeauthor{lawler14}\ resulted in only 10 useful transitions.
Nine of these lines were illustrated in Figure~7 of that paper, revealing   
that nearly all of them have depths in that star much less than 10\%.       
Scaling those depths by the V abundances found from the ionized species 
in our program stars suggests that the strongest \species{V}{i} lines 
probably would have depths $<$~2\%, thus rendering them undetectable in 
our spectra.

\subsubsection{Chromium}\label{crab}
The laboratory transition data for \species{Cr}{i} are taken from 
\cite{sobeck07} and those for \species{Cr}{ii} are from \cite{lawler17}.
The abundances derived from the ionized species are based on at least 50 lines
for each star, and there is excellent line-to-line agreement 
($\sigma$~$\simeq$~0.06, Table~\ref{tab-abmeans}).
For \species{Cr}{i} there 
are
fewer detectable lines (10--13), the scatter
among them is somewhat larger ($\sigma$~$\simeq$~0.11), and most importantly
the neutral-species mean abundances for the three stars are about 0.2~dex 
lower than those from the ionized species.
This is an issue that has been discussed previously.
\cite{kobayashi06} summarized the observational results at that time in their
Figures 20 and 21, which suggested that, at [Fe/H]~$\sim$~$-$3, abundances from
\species{Cr}{i} lines on average were $\sim$0.4~dex lower than those from
\species{Cr}{ii} lines.
Subsequent studies, \eg\ \cite{roederer14}, have not relieved this problem.

For our stars on average the abundances 
derived from the neutral species are only 
$\simeq$~0.15~dex lower than those derived from  the ionized species.
We note that nearly half of the \species{Cr}{i} transitions used here
arise from the ground state (Table~\ref{tab-lines}), and they yield abundances 
about 0.1~dex lower than those from excited states ($\chi$~$\simeq$ 1.0~eV).  
Use of only the excited-state lines would bring reasonable agreement between
\species{Cr}{i} and \species{Cr}{ii} abundances.
The NLTE computations of \cite{bergemann10} suggest that LTE abundances derived
from the neutral species are too low in metal-poor stars while those derived
from the ionized species are approximately correct.
We concur, but lack the detailed information necessary to understand the 
excitation-state dependence noted here.
Therefore we retain both Cr species as abundance indicators for our stars.
A new NLTE investigation of this element would be welcome.

\textbf{Finally, our Cr abundances are in excellent accord with those of 
\cite{lai08}: they reported [\species{Cr}{i}/Fe]~=~$-$0.07 and 
[\species{Cr}{ii}/Fe]~=~$+$0.05, while we derive $-$0.01 and $+$0.08, 
respectively.
In total, for the 11 stars in their survey with both neutral and ion
Cr abundances \citeauthor{lai08} find a larger mean difference, 
$\langle$[\species{Cr}{i}/Fe]-[\species{Cr}{ii}/Fe]$\rangle$~= $-$0.20,
but most of those stars are red giants, not main-sequence turnoff stars.}

\subsubsection{Manganese}\label{mnab}
Transition probabilities and hyperfine structure components for both
\species{Mn}{i} and \species{Mn}{ii} were taken from \cite{denhartog11}.
Abundances for our stars derived from the ionized lines have low line-to-line
and star-to-star scatter.
For the neutral species only 3--4 transitions were detectable, of which three
are the strong resonance lines at 4030.8, 4033.1, and 4034.5~\AA.~
Studies of metal-poor stars routinely report significantly deficient Mn
abundances, often [Mn/Fe] $\lesssim$~$-$0.5 (\eg, \citealt{cayrel04},
\citealt{cohen04}, \citealt{yong13}, \citealt{roederer14}).
\cite{bergemann08} used detailed computations of \species{Mn}{i} line
formation to argue that actual Mn deficiencies are only $\sim$$-$0.2~dex
if departures from LTE are taken into account for low metallicity stars.
However, their calculations indicated much larger NLTE corrections for the 
resonance lines, about $+$0.5~dex in total.

The contrast in LTE abundances between \species{Mn}{i} resonance and
higher excitation lines was discussed in SN16 and illustrated in the
lower panel of their Figure~6.
The offset is consistent with that predicted by \cite{bergemann08}.
The abundance derived from  \species{Mn}{i} lines in our program stars 
is $\simeq$~0.2~dex 
lower than that for \species{Mn}{ii} (Table~\ref{tab-abmeans}).
Among the 16 neutral-species non-resonance lines used by SN16 for \hdeight, 
we can detect just two of them (3569.5 and 4041.4~\AA; 
see  Table~\ref{tab-lines})
in our stars.
These lines yield mean abundances of log~$\epsilon$~= 2.39 in \bdzero, 2.39
in \bdone, and 2.24 in \cdthree.
These values agree well with the \species{Mn}{ii} mean abundances given in
Table~\ref{tab-abmeans}, with the caution that the 3569 and 4041~\AA\ lines
are only few percent deep in our stars.

Our results for the
\species{Mn}{i} resonance lines, based entirely on 
LTE computations, 
have not been
adjusted for suspected NLTE effects.
We believe that the computations of \cite{bergemann08} are probably correct
for the extra corrections needed for the resonance lines.
Here we have chosen simply to accept the abundances from the higher-excitation
\species{Mn}{i} transitions; see \S\ref{finalabs}.
This issue should be explored again in a future NLTE analysis.

\subsubsection{Iron}\label{feab}
Paper~1 explored Fe abundances in detail.
Here we adopt its results for \species{Fe}{i}.
More recently, \cite{denhartog19} have determined new transition probabilities
for \species{Fe}{ii}, 122 of them in the UV ($\lambda$~$<$~3300~\AA) and
10 of them at longer wavelengths.
See that paper for an extended discussion of these new $gf$ values compared
to those enumerated in the  NIST ASD \citep{kramida18}\footnote{
Atomic Spectra Database of the National Institute of Standards and Technology;
https://physics.nist.gov/PhysRefData/ASD/lines\_form.html} and in the
lab/solar empirical values for optical lines recommended by \cite{melendez09}.
In general, the \citeauthor{denhartog19}\ 
transition probabilities for blue multiplets of Fe II are lower
by 0.0--0.1~dex than those from NIST, but they are in general agreement 
with those of \citeauthor{melendez09}.
The values adopted in Paper~1 were from NIST.~
We used the \cite{denhartog19}  and \cite{melendez09} gf-values for determination of ionized-species abundances in our program stars
Our new \species{Fe}{ii} abundances are higher by 0.15~dex than reported in
Paper~1, creating an average small ion-neutral abundance offset of $+$0.1~dex.

\subsubsection{Cobalt}\label{coab}
\begin{figure}
\epsscale{1.0}
\plotone{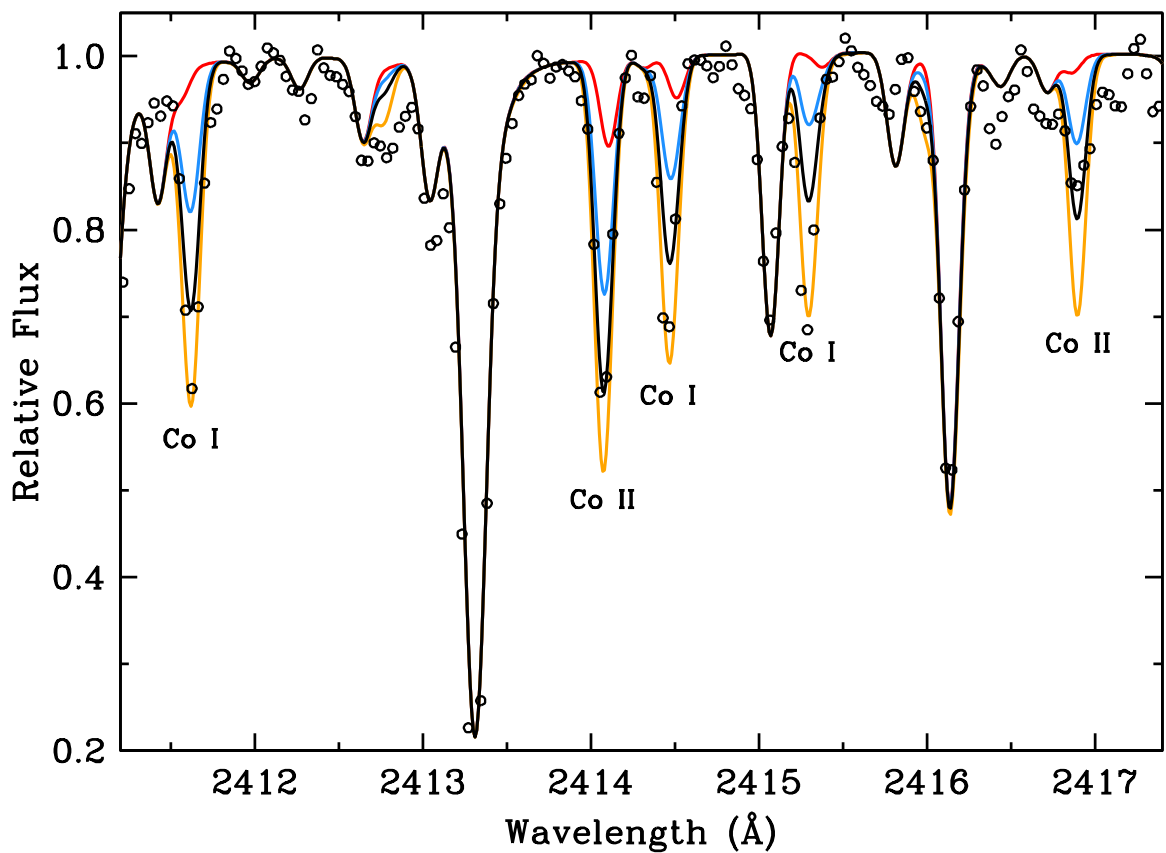}
\caption{
\label{fig2}\footnotesize
Observed spectra (open circles) and syntheses (lines) in a small spectral
region with both for \species{Co}{i} and \species{Co}{ii} transitions.
For the syntheses, the red color indicates what the spectrum would look
like with no Co contribution and the black color is for the mean Co abundance
from \species{Co}{ii} features. 
The blue and orange lines represent the \species{Co}{ii}-based abundance
decreased 
and increased by 0.4~dex.
}
\end{figure}

Both \species{Co}{i} and \species{Co}{ii} have recent lab transition
probability and hyperfine structure analyses (\citealt{lawler15},
\citealt{lawler18}).
But application of these line data to the three program stars produces the
most significant neutral/ion abundance clash of this study.
The mean values, from Table~\ref{tab-abmeans}, are 
$\langle$[\species{Co}{i}/H]$\rangle$~=~$-$2.53 while
$\langle$[\species{Co}{ii}/H]$\rangle$~=~$-$2.93, smaller by a factor of 2.5.
In Figure~\ref{fig2} we illustrate the problem with a comparison of
observed and synthetic spectra of \bdzero\ in a small UV region that
contains several \species{Co}{i} and \species{Co}{ii} lines.
Inspection of this figure easily suggests that the abundances from
neutral-species transitions are substantially larger than those from
ionized-species ones.
The \species{Co}{i} abundances for our stars are based on 28--36 lines,
an order of magnitude more than most large-sample surveys.
The \species{Co}{ii} abundances are based on 16--21 lines, none of which
have played significant roles in past Co studies except SN16.

Six elements (Ti, Cr, Mn, Fe, Co, and Ni) have abundances derived from
both neutral and ionized species.
Inspection of Table~\ref{tab-abmeans} and Figure~\ref{fig1} shows that
for five of these elements the species abundances agree with each other to
within $\sim$~0.2~dex, sometimes much better than that.
The glaring exception in our stars is Co, for which
$\langle$log~$\epsilon$(\species{Co}{i})$-$log~$\epsilon$(\species{Co}{ii})$\rangle$
$\simeq$~$+$0.4.

Co has a prominent place in abundance studies of very metal-poor stars.
For nearly three decades spectroscopic surveys have reported that for
\textbf{metallicities in the range $-$2.5~$\lesssim$~[Fe/H]~$\lesssim$~$+$0.0 
Co retains}
its solar abundance ratio, [Co/Fe]~$\simeq$~0.0, but for lower
metallicities the Co relative abundance steadily (perhaps) increases,
reaching [Co/Fe]~$\sim$~$+$0.5 at [Fe/H]~$\sim$~$-$3.0 (\eg,
\citealt{ryan91,ryan96}, \citealt{mcwilliam95}, \citealt{cayrel04},
\citealt{cohen04}, \citealt{barklem05}, \citealt{yong13}, 
\citealt{roederer14}).
Nearly all metal-poor star surveys have used only \species{Co}{i} transitions,
as \species{Co}{ii} features only become strong in the UV.~
Typically only a handful of \species{Co}{i} lines have contributed to the
reported Co abundances in these papers.
Additionally, NLTE computations by \cite{bergemann10} suggested that there
may be large positive abundance corrections for \species{Co}{i} lines in
very metal-poor stars, +0.5~dex or more.
Their Figure~6 proposes that [Co/Fe]$_{\rm NLTE}$~$\sim$~$+$0.7 for
[Fe/H]~$<$~$-$2.

If we had studied only \species{Co}{i} in these stars, we would have
confirmed the high Co from past investigations of very low metallicity 
stars.
But neutral Co is just a trace species of Co in very metal-poor
main-sequence turnoff stars.
For \hdeight, Figure~2 of SN16 suggests that
$N_{\rm Co II}$/$N_{\rm Co I}$~$\gtrsim$~25, meaning
that more than about 95\% of the element exists as singly-ionized Co.
\cite{bergemann10} do not consider NLTE line formation for 
\species{Co}{ii}
in detail, but they suggest that the corrections to the LTE abundances in our
types of stars will be small.
Acceptance of the ionized-species abundance puts the relative [X/H]
values for Co in agreement with those of its Fe-group element neighbors;
i.e., the mean of [Co/Fe] in the target stars is approximately solar.
We believe that the large abundances indicated by \species{Co}{i}
are not correct.

\subsubsection{Nickel}\label{niab}
\cite{wood14b} presented laboratory transition probabilities\footnote{
Additionally, this paper considered isotopic wavelength splitting from
the five stable Ni isotopes.
However, isotopic shifts for \species{Ni}{i} only become large enough to
materially affect line shapes in the red spectral region, 
$\lambda$~$>$~6000~\AA.
This line complexity is negligible in our stars, which have no detectable
transitions in the red.}  
for 371 \species{Ni}{i} transitions.
With these new $gf$ values, we obtained consistent results for our stars
with a mean abundance log~$\epsilon$~= 3.40, using 35--45 lines per
star, with line-to-line scatters $\sigma$~$\simeq$~0.08 
(Table~\ref{tab-abmeans}).
Happily, the derived abundances from \species{Ni}{ii} transitions is 
essentially in agreement, with a mean log~$\epsilon$~= 3.36
in the three program stars.
However, some cautions should be raised here:\
\textit{(a)} only 7--8 ionized-species lines could be detected in our stars;
\textit{(b)} all of these lines are deep in the UV, $\lambda$~$<$~2500~\AA;
\textit{(c)} the abundance scatters are large, $\sigma$~$\simeq$~0.16; and
\textit{(d)} perhaps most importantly, \species{Ni}{ii} laboratory $gf$
values have not been revisited recently.
\cite{wood14b} adopted the transition probabilities of \cite{fedchak99}
and they have been used in our analysis.
Since Ni is the heaviest Fe-group element with detectable neutral- and
ionized-species transitions in very metal-poor stars, a new laboratory
\species{Ni}{ii} lab study would be worthwhile.
 
\subsubsection{Copper}\label{cuab}
The available transitions are severely limited here.
All strong \species{Cu}{ii} lines occur well shortward of our 
2300~\AA\ \textit{HST}/STIS wavelength limit.
Of the possible \species{Cu}{i} transitions, we were only able to work with
the resonance lines at 3247.5 and 3273.9~\AA. 
The transitions at 5105.5 and 5782.1~\AA, used in most stellar Cu abundance 
studies, are far too weak for detection in our stars.
From the resonance lines in the three program stars, 
we derived a mean
abundance of $\langle$log~$\epsilon\rangle$~$\simeq$~0.6, or 
$\langle$[Cu/H]$\rangle$~$\simeq$~$-$3.5.
However, it is possible that these very low abundances may be severe 
underestimates, as some studies (\eg, \citealt{andrievsky18b}, \citealt{shi18}) 
have argued that LTE-based abundances should be corrected upward by large 
factors, typically by at least $+$0.5~dex in very metal-poor stars.

Some information on stellar abundances from \species{Cu}{ii} now is available.
For two metal-poor ([Fe/H]~$\simeq$~$-$2.4) red giants. \cite{roederer14b} 
employed four \species{Cu}{ii} lines in the 2030--2130~\AA\ spectral range 
to determine [Cu/H]~$\simeq$~$-$2.8 .
More directly, for \hdeight\ \cite{roederer18a} used these same 
\species{Cu}{ii} features, deriving [Cu/H]~$=-$2.75 .  
\textbf{We have added their \species{Cu}{ii} abundance to the bottom panel of
Figure~\ref{fig1}.}
The abundances from neutral lines for this star are [Cu/H]~$\simeq -$3.2
(SN16) and [Cu/H]~$\simeq-$3.1 (\citeauthor{roederer18a}).
The 0.3--0.4~dex larger abundance from \species{Cu}{ii} lines suggests that
NLTE corrections of this magnitude probably should be applied to our
\species{Cu}{i} results.
However, we lack ionized-line abundances in our program stars, which are
significantly more metal-poor than those studied by 
\cite{roederer14b} and \citeauthor{roederer18a}, 
and our spectra do not cover the shorter UV wavelengths 
where the Cu~\textsc{ii} lines are found.
Therefore we have chosen here not to make arbitrary adjustments to our 
LTE-based abundances, but just as in the cases of \species{Cr}{i} and 
\species{Mn}{i} we urge a comprehensive NLTE study of \species{Cu}{i} 
and \species{Cu}{ii} for our program stars.

\subsubsection{Zinc}\label{znab}
\species{Zn}{ii} features are found at UV wavelengths shorter than 
those covered by our STIS spectra.
(The resonance lines of this species occur at 2025.5 and 2062.0~\AA.)~
As in the case of Cu, we can only detect the \species{Zn}{i} resonance lines
at 3302.6 and 3345.0~\AA; all higher-excitation lines of the 
neutral species are absent in our spectra.
The 3302 and 3345~\AA\ lines are extremely weak, and in fact they cannot be
detected in \cdthree.

Support for the \species{Zn}{i} results comes from the 
literature studies discussed for Cu in \S\ref{cuab}.
The \species{Zn}{i} and \textsc{ii} lines 
in the two metal-poor red giants studied by
\cite{roederer14b} yielded overabundances 
[Zn/Fe]~$\simeq$~$+$0.3, comparable to 
those of our program stars.
The \cite{roederer18a} abundances for neutral and ionized Zn lines in
\hdeight\ are [Zn/Fe]~=~$+$0.23 and $+$0.13, respectively, 
while SN16 reported [Zn/Fe]~=~$+$0.16 from just the neutral lines
in this star.
\textbf{Their abundance for \species{Zn}{ii} has been 
entered into the bottom panel of Figure~\ref{fig1}.}
This limited sample of \species{Zn}{ii} abundances 
suggests that
\species{Zn}{i} features yield reliable abundances in this metallicity regime.
[Zn/Fe] overabundances appear to be real.

\section{FINAL ABUNDANCES FOR THREE VERY LOW METALLICITY STARS}\label{finalabs}

The Fe-group relative abundances are very similar in our three main-sequence
turnoff stars with metallicities [Fe/H]~$\simeq$~$-$2.9.  
This is apparent from inspection of the log~$\epsilon$ and [X/H] values
in Table~\ref{tab-abmeans}.

\begin{figure}
\epsscale{1.0}
\plotone{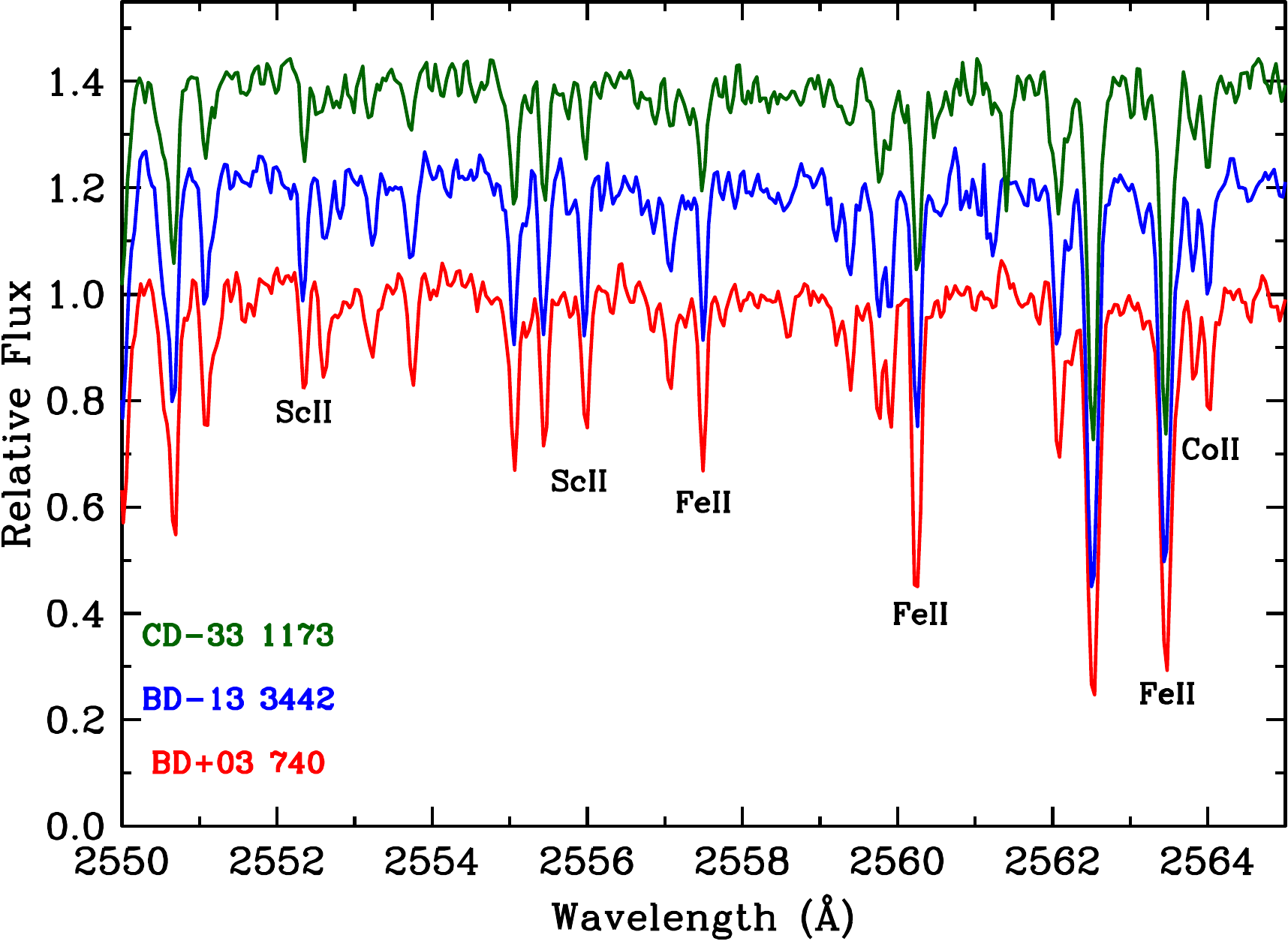}
\caption{
\label{fig3} \footnotesize
Spectra of the three program stars in a small UV region.
Essentially all of the absorption features can be accounted for by
Fe-group transitions, some of which are labeled in the figure.
For clarity the relative flux scales for \bdone\ and \cdthree\ have been
shifted by $+$0.2 and $+$0.4, respectively.
}
\end{figure}

Unsurprisingly this abundance similarity extends to the [X/Fe] values that are
shown in Table~\ref{tab-abratios}.
\textbf{These numbers are computed from the [X/H] and [Fe/H] values in 
Table~\ref{tab-abmeans}.}
The star-to-star variations in [\species{X}{i}/Fe] and [\species{X}{ii}/Fe]
are usually $\lesssim$~0.1~dex.
The agreement among the three stars can also be seen by inspection of
small spectral regions, as we illustrate in Figure~\ref{fig3}.
In this figure the line strengths for \cdthree\ are weaker than those of
the other two stars, due to its lower metallicity and substantially higher
\teff.
But the ratios of depths of the \species{Sc}{ii} and \species{Co}{ii}
lines compared to those of \species{Fe}{ii} does not change much among
the stars, nor do the line depths of other species not shown in this
figure.
Therefore we also have tabulated the three-star means of these values
in this table; they serve as good representations for our final species
abundances of our program stars.

For the elements Cr, Mn, and Cu, Table~\ref{tab-abratios} has double entries
for the neutral species abundances.  
As discussed in \S\ref{abspecies}, both observational and 
theoretical (NLTE) evidence for Cr and Mn 
strongly suggests that their $\chi$~=~0~eV 
resonance lines are not well represented by LTE computations for 
metal-poor stars.
LTE abundances from higher-excitation neutral species transitions are more
in accord with the abundances from the ionized species for these elements.
Therefore the Table~\ref{tab-abratios} labeled ``Cr-rev'' and ``Mn-rev'' 
contain the neutral-species abundances without inclusion of their
resonance lines.

For Cu, NLTE studies agree that major upward abundance corrections are
needed.
Unfortunately, for Cu in our stars we have neither high-excitation 
\species{Cu}{i} nor \species{Cu}{ii} transitions to provide observational 
guidance.
Therefore for the neutral-species abundances in the row labeled ``Cu-rev''
we have applied an offset of $+$0.50~dex to the observed Cu abundances.
We emphasize that this shift is an arbitrary single value.  
It probably is approximately correct, but it should be viewed with much
caution.

\begin{figure}
\epsscale{1.0}
\plotone{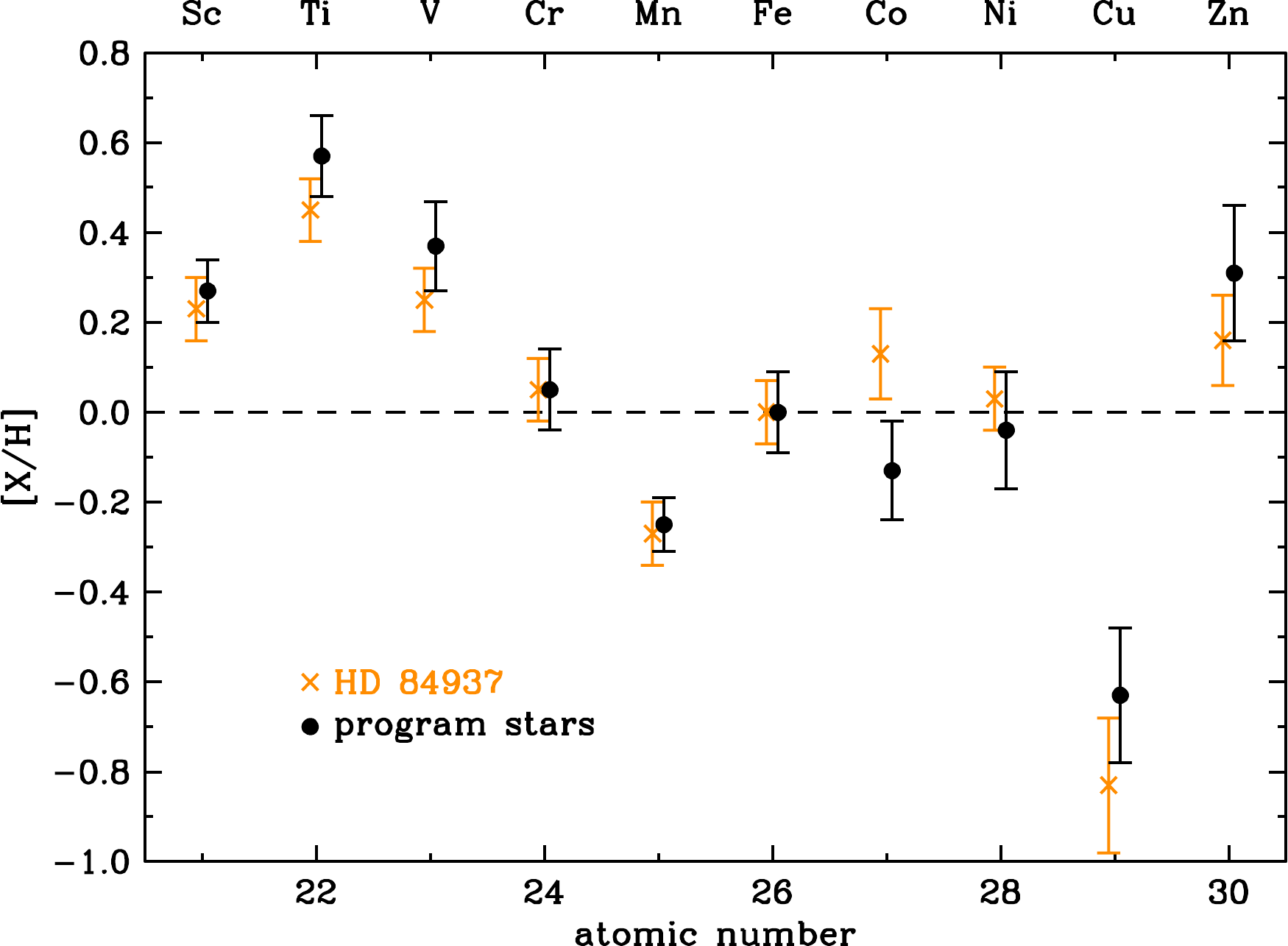}
\caption{
\label{fig4} \footnotesize
Final elemental relative abundance ratios for the three program stars.
See \S\ref{finalabs} for explanations \textbf {of the chosen ratios and the
associated error bars.}.
The relative abundance ratios for \hdeight\ (SN16) are also shown
in the figure.
}
\end{figure}

In Figure~\ref{fig4} we display our final elemental abundance ratios
\textbf{for our target stars. Most of the illustrated values are  
the means (ions and neutrals, where they both have been measured) 
taken from the 
last column of Table 4, except for cobalt where we have opted to use
Co II. 
Also illustrated are abundance determinations for \hdeight\ from SN16.}
\textbf{Except for Cu and Zn the error bars in this 
figure represent the means of  
the scatters (sigmas). More definitive values for these error bars 
will only be possible with additional NLTE studies.
For Cu and Zn there are not enough points to assess 
realistic scatter uncertainties, so we have assigned $\pm$ 0.15 
to them. 
(This is 
consistent with the few sigma values obtained for these two
elements,  as shown in Table 3.)}
\textbf{The most obvious} 
departure from Solar abundances is the coordinated
large overabundant element set of Sc, Ti, and V.~
In the next section we argue that this Fe-group abundance signature 
exists in larger spectroscopic surveys as well.

\section{Fe-GROUP ABUNDANCE RATIOS IN THE TARGET STARS AND OTHER
            LOW-METALLICITY STARS\label{results}}

The main results of our Fe-group abundance study of three very metal-poor
main sequence turnoff stars are as follows:\
(a) the [X/Fe] values agree within the errors for all species in all stars;
(b) the lightest Fe-group elements are clearly all overabundant, with 
$\langle$[X/Fe]$\rangle$~$\sim$~+0.4; and 
(c) the abundances derived from \species{Co}{i} lines are uniformily 
larger than those from \species{Co}{ii} by about 0.4~dex.
Among the other elements, we find that [Cr/Fe]~$\simeq$~[Ni/Fe]~$\simeq$ 0.0,
and [Mn/Fe]~$\simeq$~[Cu/Fe]~$\simeq$ $-$0.2 after estimated NLTE corrections
are applied (\S\ref{mnab} and \S\ref{cuab}).
Our Zn abundance is based on two very weak lines detected in \bdzero\ and
\bdone\ but not in \cdthree; little weight should be given to our results for
this element.

\begin{figure}                                                
\epsscale{1.0}                                                
\plotone{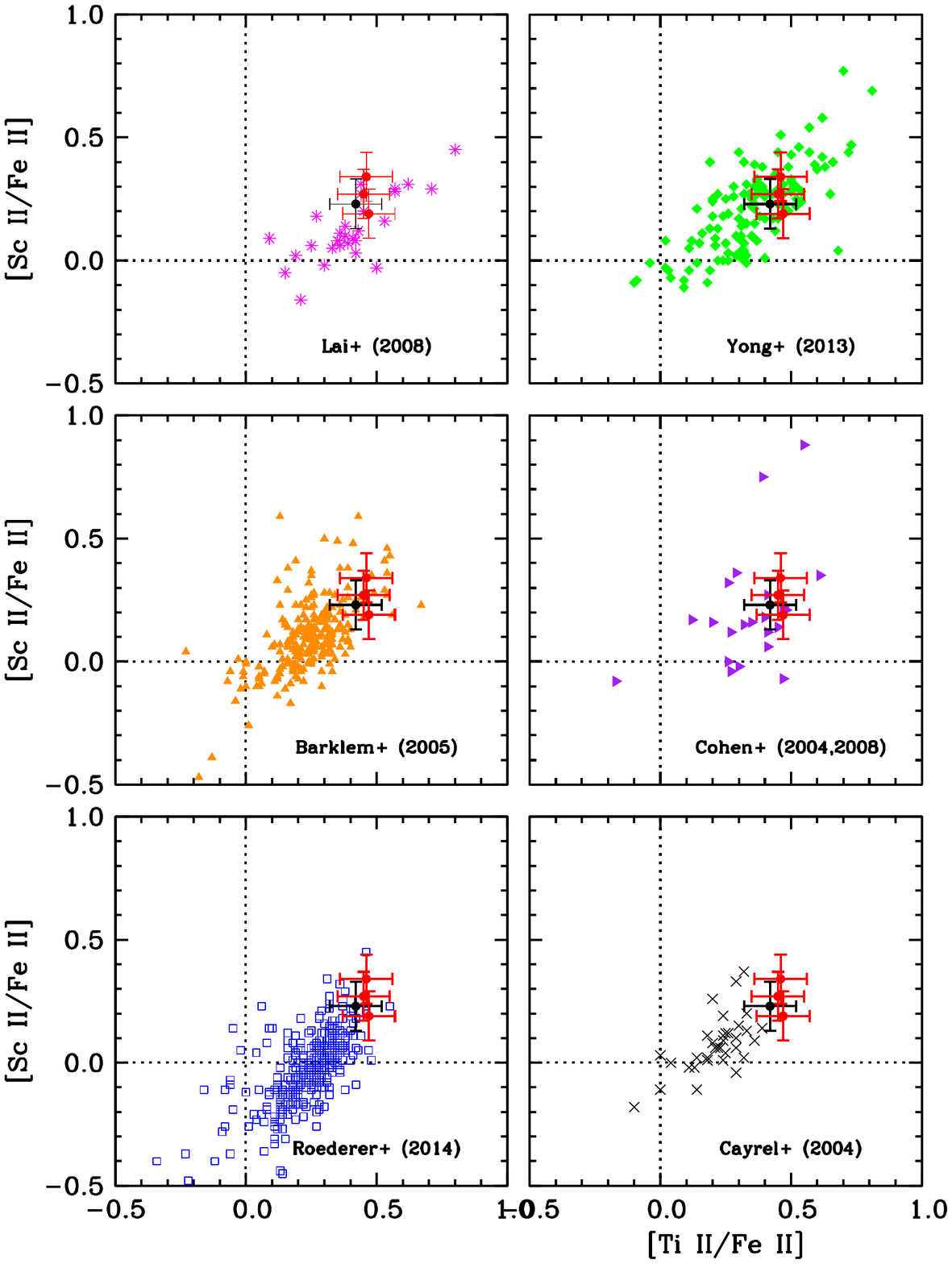}                                           
\caption{                                                     
\label{fig5} \footnotesize                                   
Correlations of [Sc/Fe] and [Ti/Fe] ratios in six major halo-star
abundance surveys, \cite{cayrel04}, \cite{cohen04,cohen08}, \cite{barklem05},
\cite{lai08}, \cite{yong13}, and \cite{roederer14}.
In each panel the \hdeight\ value from SN16 is shown as a black dot with 
error bars, and the three program stars are red dots with error bars.
}                                                             
\end{figure}

SN16 discovered coordinated  relative abundances of Sc, Ti, and V in \hdeight.
They suggested that these abundance correlations generally can be found
among results from surveys of halo stars with [Fe/H]~$\lesssim$~$-$2.
With even larger Sc-Ti-V relative overabundances found in our program 
stars, we revisit this issue here.
In Figure~\ref{fig5} we plot 
[\species{Sc}{ii}/\species{Fe}{ii}] versus 
[\species{Ti}{ii}/\species{Fe}{ii}] 
from six major studies along with our results.
Different panels of this figure contain data from \cite{cayrel04},
\cite{cohen04,cohen08}, \cite{barklem05}, \cite{lai08}, \cite{yong13}, 
and \cite{roederer14}.
All data sets show positive correlations between Sc and Ti abundances,
and our program stars fall within the general set of abundances in each
survey shown in this figure.

The large surveys included in Figure~\ref{fig5} and figures to follow
are based on optical data, and their spectra have variable 
$S/N$ values.
Thus their abundances usually are based on substantially fewer lines 
than those of our program stars and \hdeight.  
In particular, \cite{barklem05} gathered ``snapshot'' high-resolution spectra,
with the primary goal of identifying \rpro-rich metal-poor giants.
They chose to use relatively short exposure times in order to
observe a large set of stars (373).
Their spectra had modest fluxes, with $\langle S/N\rangle$~$\sim$~50.
This led \citeauthor{barklem05}\ to report fewer 
abundances 
of elements only represented by a 
small number of
weak transitions (\eg, V) than those with a rich set of lines (\eg, Ti).
Therefore we chose to eliminate stars for which they obtained spectra with 
$S/N$~$<$~35, and those for which they derived abundances of Sc and Ti 
but not V.~
These two choices combined to produce the \citeauthor{barklem05}\ sample with
the highest probability of good abundances for this element trio.
Additionally, the panel Figure~\ref{fig5} with \cite{cohen04,cohen08} 
results do not show the correlated trends seen in other surveys.  
However, the earlier \citeauthor{cohen04}\ paper included only main
sequence stars, and the later \citeauthor{cohen08}\
paper concentrated on stars with 
[Fe/H]~$\lesssim$~$-$3.4, making it difficult to measure many \species{Sc}{ii} 
and especially \species{V}{ii} lines.
We regard the \citeauthor{cohen04}\ abundances to be consistent with those
from the other surveys, which have usually more favorable line detection
probabilities for these elements.

\begin{figure}
\epsscale{1.2}
\plotone{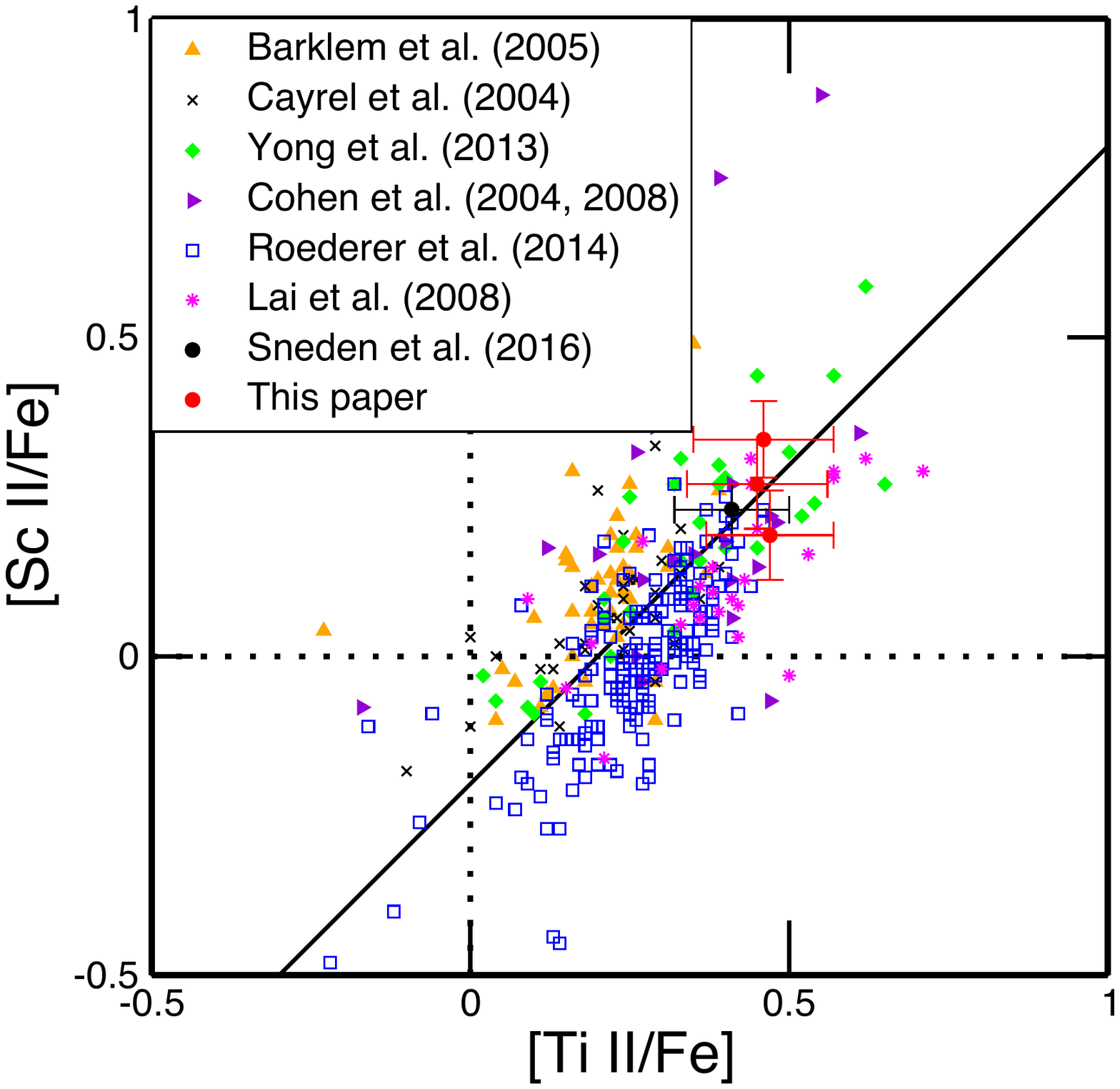}
\caption{
\label{fig6}\footnotesize
Abundance ratios [Sc/Fe] versus [Ti/Fe] from ionized transitions of each
element.
The open squares are from \cite{roederer14},  
filled diamonds from \cite{yong13}, 
filled right-facing triangles from \cite{cohen04,cohen08}, 
filled upward-facing triangles from \cite{barklem05}, 
x's are from \cite{cayrel04}, stars from \cite{lai08}
and the filled circles  for
\bdzero, \bdone, \cdthree, and \hdeight\ derived in this paper or in SN16.
The horizontal and vertical (dotted) lines denote the solar abundance ratios
of each element.
The solid line represents a 45$^{\circ}$ slope.
}
\end{figure}

In Figure~\ref{fig6} we merge all of the data sets into an overall
[Sc/Fe] vs. [Ti/Fe] correlation plot.
Also shown for illustrative purposes is a solid black line at a 45$^{\circ}$
angle, placed to go through the mean of our [Sc/Fe] values for \hdeight,
\bdzero, \bdone, and \cdthree.
All of the individual data sets, obtained with different atmospheric models,
atomic line sets, and abundance techniques, cluster along this line, 
indicating a strong abundance correlation between these two elements.
We have chosen not to apply additive constants to individual survey results
in order to tighten the overall trend, because that would not materially
change what is clear in this figure.
That is, [Sc/Fe] and [Ti/Fe] vary in concert over a range of $\sim$~0.5~dex in
individual very metal-poor ([Fe/H]~$\lesssim$~$-$2) Galactic halo stars.
Abundances for our program objects lie at the high end of these abundance 
ratios, but other surveys have found stars with even higher values.

\begin{figure}
\epsscale{1.2}
\plotone{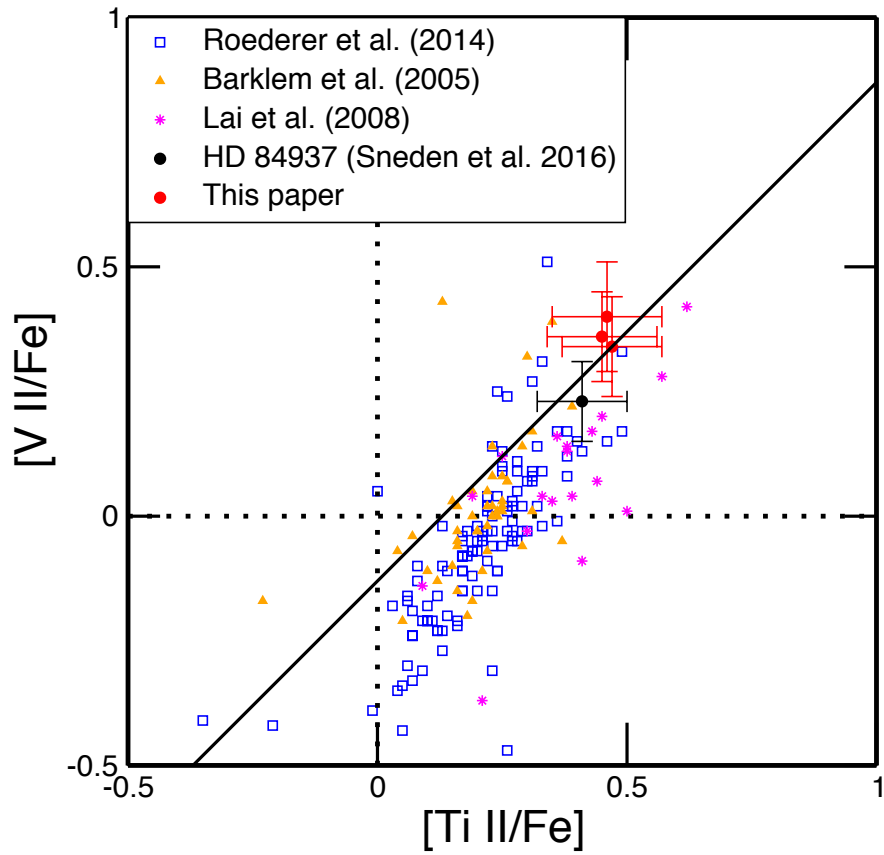}
\caption{
\label{fig7}\footnotesize
Abundance ratios [V/Fe] versus [Ti/Fe] from ionized transitions of each
element.
The symbols are as in 
Figure~\ref{fig6}.
}
\end{figure}

We make a similar abundance comparison with [V/Fe] and [Ti/Fe] in 
Figure~\ref{fig7}.
As in Figure~\ref{fig6} our results are again shown as filled circles,
and again we have placed the 45$^{\circ}$ trend line to pass through the 
mean values of our program stars and \hdeight.
The other data sets are from \cite{barklem05}, \cite{lai08}, and
\cite{roederer14}.
The surveys of \cite{cohen04,cohen08} and \cite{yong13} did not report
abundances from either \species{V}{i} or \species{V}{ii} lines, 
due to the general
weakness of this element's transitions in the optical spectral region.
Indeed, the abundances from \species{V}{ii} lines in \citeauthor{barklem05}\
are based only
on the line at 3951.9~\AA, and 
\citeauthor{roederer14}\ used only that line and
one other at 4005.7~\AA.~
Only the \citeauthor{lai08}\ study 
used more \species{V}{ii} lines, most found at
blue wavelengths down to about 3500~\AA.~
Therefore, we  do not regard the $\sim$0.1~dex offset between our abundances
and those of other literature sources as significant.
Clearly there is a direct correlation between [V/Fe] and [Ti/Fe] among all 
of the data sets shown in Figure~\ref{fig7}. 
The three lightest Fe-group elements, Sc, Ti and V, are correlated in 
metal-poor halo stars, and thus appear to have a closely related 
nucleosynthetic origin.

\begin{figure}
\epsscale{1.2}
\plotone{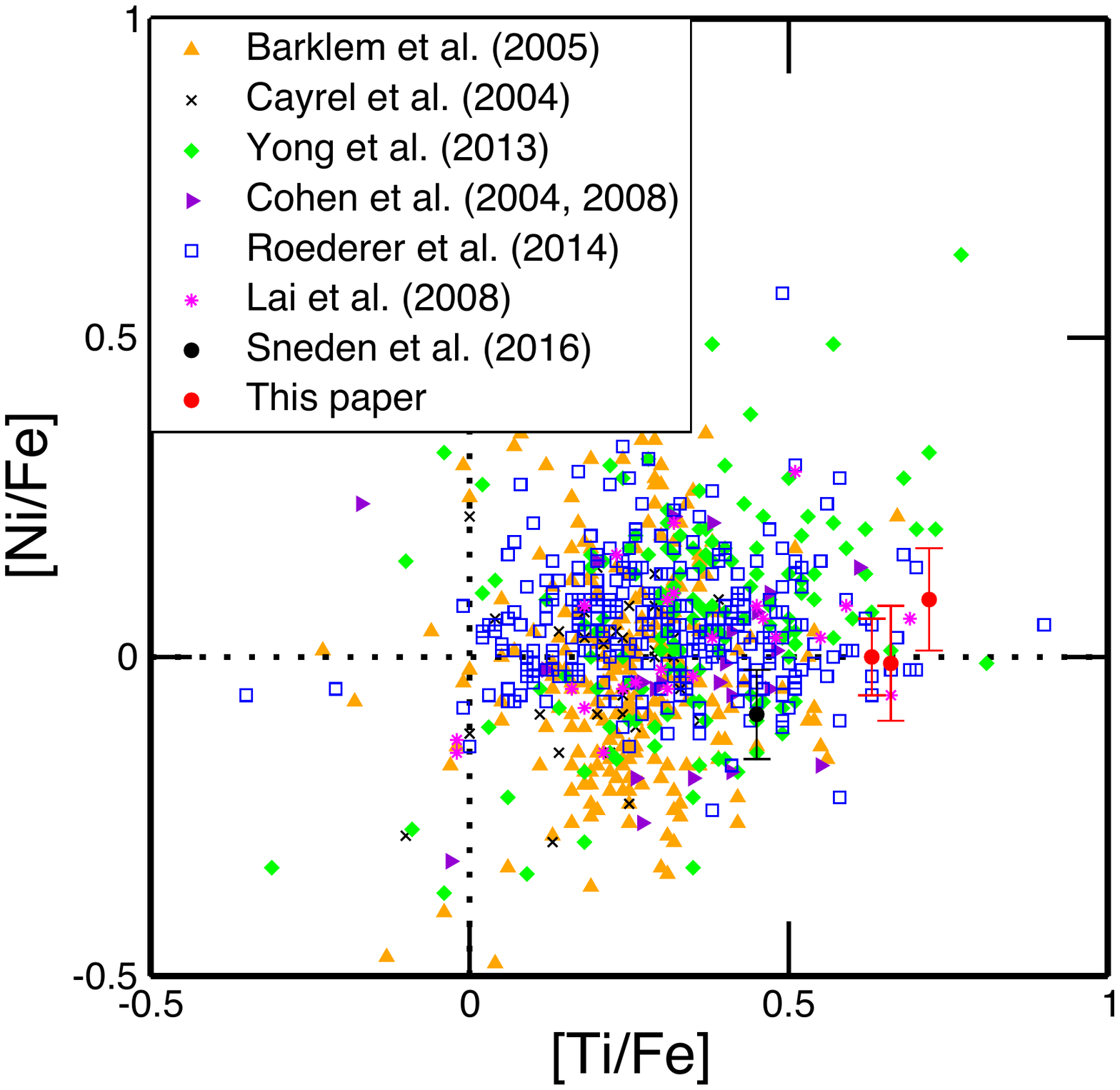}
\caption{
\label{fig8}\footnotesize
Abundance ratios [Ni/Fe] versus [Ti/Fe] from neutral  transitions of each
element.
The symbols are as in 
Figure~\ref{fig6}.
}
\end{figure}

As a check on whether this correlation extends to other iron-peak elements,
we examine [Ti/Fe] versus [Ni/Fe], as illustrated in 
Figure~\ref{fig8}.
A similar comparison was made in SN16, albeit
with more limited data. 
We again include the data 
sets of \cite{cayrel04}, \cite{barklem05}, \cite{cohen04,cohen08}, 
\cite{lai08}, \cite{yong13}, and \cite{roederer14}, 
along with our new precise abundance values for low-metallicity stars. 
Figure~\ref{fig8} shows that Ti abundances are not correlated with
Ni abundances, unlike the case of correlated Sc, Ti, and V abundances.

\subsection{Nucleosynthesis Origins of the Fe-group elements\label{nucleo}}

The production of Ti and V in core-collapse supernovae (CCSNe)
has been studied
in detail 
(see, e.g., discussion and references in \citealt{woosley95}).
Explosive stellar yields are also available for different metallicities
\cite[e.g.,][]{woosley95,limongi00,rauscher02,heger10,chieffi13,nomoto13,pignatari16,curtis19}.
These iron-group elements are synthesized in the inner layers of stars 
undergoing complete or 
\textbf{incomplete silicon burning.}
The production of these elements is tied to the explosion energies 
and details of the mass cut, which determines how much material is ejected in
the explosion.
Thus, in CCSNe ejecta, Ti and V
are synthesized in the
complete Si burning regions, but both elements 
tend to be underproduced relative to the observed abundance data 
for metal-poor halo stars.
\textbf{\citeauthor{curtis19} discuss production of Sc, Ti and V in detail.
A couple of points from that work are worth mentioning here.
First, Sc production can be complicated in these explosions, 
but it is mostly produced as Ti in the inner layers of the explosion.
Second, although past spectroscopic results have linked Ti to $\alpha$
elements such as Mg, Si, and Ca, due to its relative overabundance,
Ti is not a true $\alpha$ element (i.e., produced by the capture of 
$\alpha$ particles leading to stable isotopes with even numbers of protons 
and neutrons; e.g., $^{20}$Ne and $^{40}$Ca).
Instead, Ti has a similar origin to Sc and V, and it is synthesized in Si 
burning regions deep in CCSNe, with the dominant Ti isotope
being $^{48}$Ti formed from the decay of $^{48}$Cr.}

In the future it will be important to make further precise abundance
determinations in the metal-poor halo stars
utilizing the new atomic laboratory data for the iron-peak elements.
X.\ Ou et al.\ (in prep.)\ present a new reanalysis of the
V abundances of nearly 300~stars in the \cite{roederer14} sample.
That study makes use of improved atomic data for \species{V}{i} 
\citep{lawler14} and \species{V}{ii} \citep{wood14a} to
expand the number of lines useful for a V abundance analysis, resulting 
in smaller abundance uncertainties and more stars with V detections.
That study supports the results of SN16 and the present work
that Sc, Ti, and V abundances are correlated in metal-poor stars.

\subsection{Implications for Early Galactic Nucleosynthesis\label{gce}}

\begin{figure}
\epsscale{1.2}
\plotone{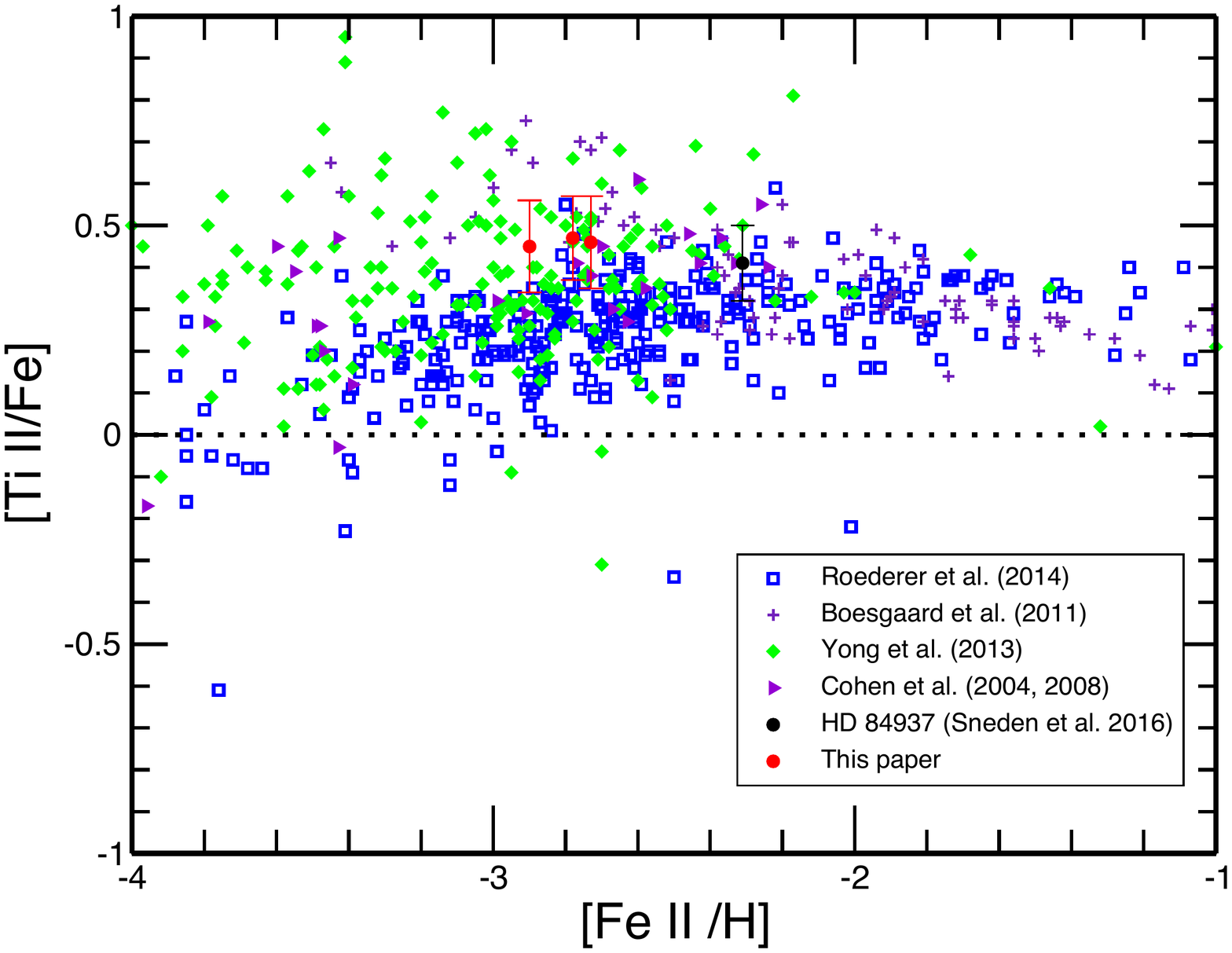}
\caption{
\label{fig9} \footnotesize
Abundance ratios [Ti~\textsc{ii}/Fe] 
plotted as function of [Fe~\textsc{ii}/H] metallicity.
The symbols are as in 
Figure~\ref{fig6} with plus signs from \cite{boesgaard11}. 
See text for detailed discussion.
}
\end{figure}

In Figure~\ref{fig9} 
we show the [Ti/Fe] abundance ratios
for our target stars
(shown as filled circles) as a function of metallicity. 
Also illustrated are the derived abundances 
from other low-metallicity star surveys:\
\cite{roederer14} (filled squares), \cite{yong13} (filled diamonds), 
\cite{boesgaard11} (plus signs), and 
\cite{cohen04,cohen08} (filled right-facing triangles).
It is clear that the values of [Ti/Fe] for our target stars are high, 
approaching $+$0.5, significantly higher than the solar value.
In fact the data from all of the surveys indicates that [Ti/Fe] 
is substantially  enhanced in low-metallicity stars.
This trend has been reported before (SN16), 
but these new precise abundance values 
based on new laboratory atomic data for
the three very low-metallicity target stars
strongly support this result.
We also note that recent work has suggested this trend continues 
at even lower metallicities,
as \cite{nordlander19} report a value of [Ti/Fe]~$= +$0.82
for a star with a metallicity of $-$6.2.

These results suggest that at early Galactic times 
the synthesis sites for iron-peak element production
make an overabundance of Ti.
While CCSNe models have difficulty producing enhanced Ti 
(see discussion in SN16),
higher explosion energies in hypernovae can produce larger amounts 
of certain iron-peak elements
\citep{umeda02,kobayashi06}.
Galactic chemical evolution models 
(e.g., \citealt{kobayashi06,kobayashi11b})
that assume a 50\% hypernova fraction
can reproduce the overall abundance trends,
but they have difficulty in achieving 
the actual abundance values seen in the data.
These comparisons suggest that an even larger proportion of 
hypernovae might occur at very low metallicities and
early in the history of the Galaxy
(cf.\ \citealt{ezzeddine19}).

\begin{figure}
\epsscale{1.2}
\plotone{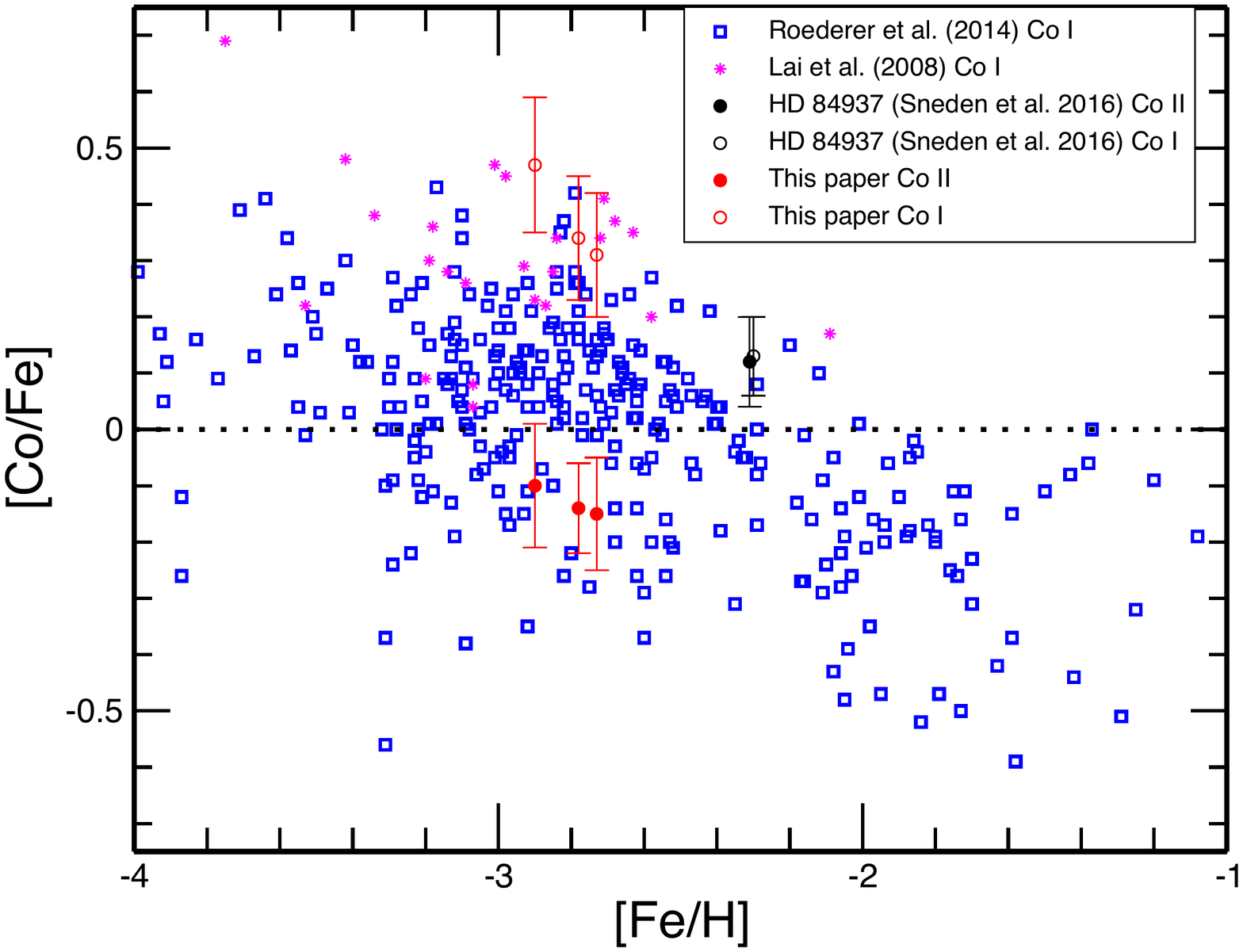}
\caption{
\label{fig10} \footnotesize
Abundance ratios [Co/Fe] plotted as function of [Fe/H] metallicity
The symbols are as in 
Figure~\ref{fig6}.
See text for detailed discussion.
}
\end{figure}

As noted above in \S\ref{abspecies}, early surveys, 
employing only neutral Co abundances, 
indicated overabundances of [Co/Fe] at low metallicities.
We examine the abundance trends of [Co/Fe] in Figure~\ref{fig10}
using data from \cite{lai08}, \cite{roederer14}, SN16, and
our new abundance determinations for the three low metallicity target stars. 
In general the data are not consistent with an overabundance at
low metallicities. 
The exceptions are the new values for our target stars when
employing abundances derived
only the neutral (minority)species,
which are significantly larger than the abundances derived from
Co~\textsc{ii} lines. 
It is also clear in the figure that both neutral and ionized species of Co
give the same result for the somewhat more metal-rich \hdeight\ (SN16).
We conclude that the  [Co/Fe]  abundance ratios derived from 
singly-ionized Co are correct,
and that differences between the neutral and ionized results are mostly likely
due to NLTE effects. 
Clearly, this issue should be examined in the future.

\begin{figure}
\epsscale{1.2}
\plotone{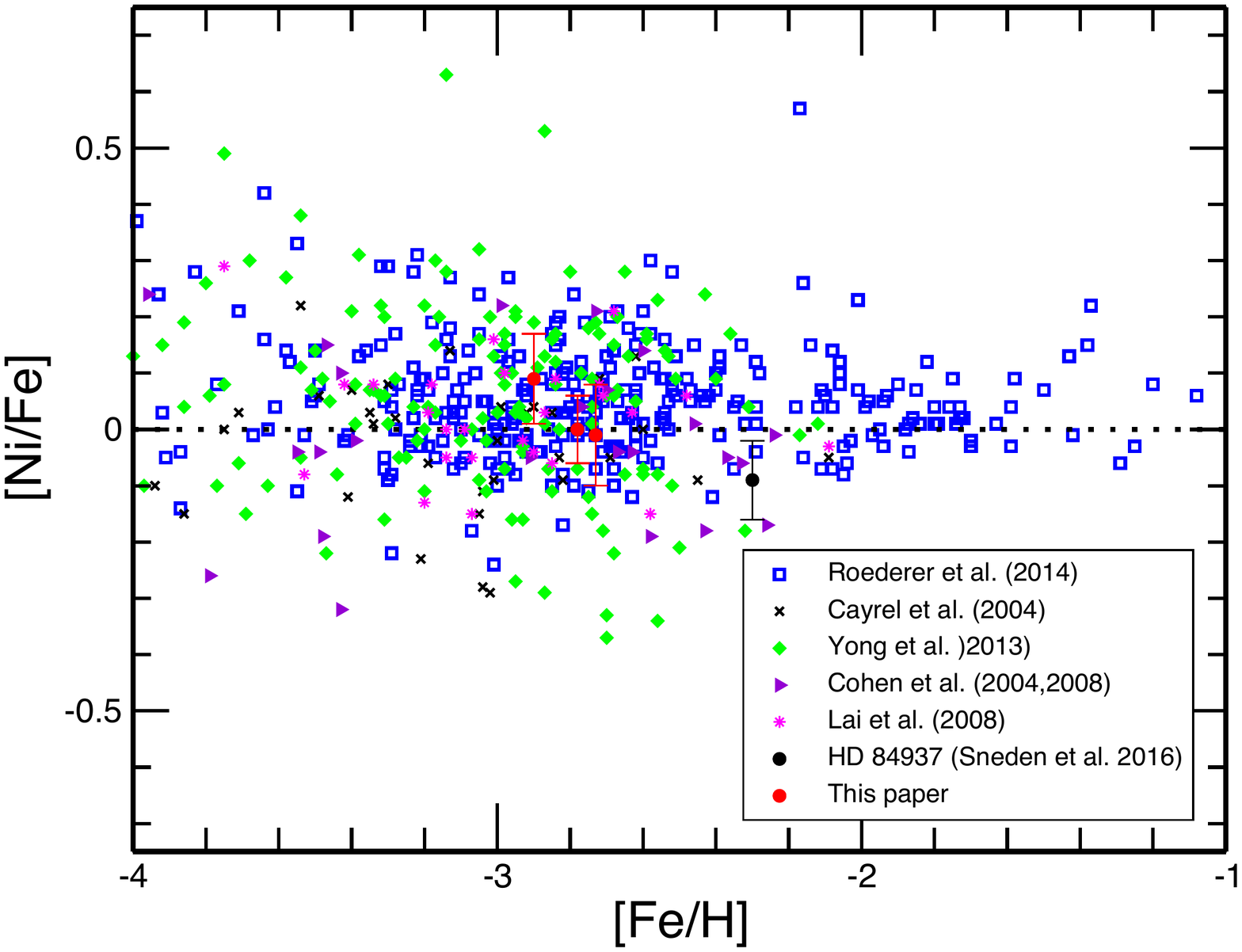}
\caption{
\label{fig11} \footnotesize
Abundance ratios [Ni/Fe] plotted as function of [Fe/H] metallicity.
The symbols are as in 
Figure~\ref{fig6}.
See text for detailed discussion.
}
\end{figure}

As a final iron-peak element abundance comparison, 
we examine the chemical evolution of [Ni/Fe] in Figure~\ref{fig11}.
Our new abundance determinations for the program stars, 
along with data from \cite{cayrel04}, 
\cite{cohen04,cohen08},  \cite{lai08}, \cite{yong13}, and \cite{roederer14}, 
indicate mostly solar values of [Ni/Fe] over a large metallicity range.

\section{SUMMARY AND CONCLUSIONS\label{sum}}

We have determined new precise abundances for the iron-peak elements 
(Sc--Zn) in three very metal-poor ([Fe/H] $\approx$ $-$3) stars:\
\bdzero, \bdone,  and \cdthree.
Our high-resolution spectroscopic data sources include \textit{HST}/STIS UV
spectra and optical ground-based spectra from several sources.
This is the first study of the complete set of Fe-group elements with 
recent, consistent lab data.

Comparing our new abundance values with other surveys, we conclude that
the three lightest Fe-group elements, Sc, Ti, and V, are correlated in
metal-poor halo stars.
Our target stars fall on the high-end of this general, relatively unstudied 
correlation.
It has been reported previously that the iron-peak elements Sc, V, and 
especially Ti are often overabundant in stars at low metallicities.
Our new abundance results for these three metal-poor halo stars strongly 
support this finding. 
These results suggest a similar astrophysical origin for these three 
elements and place strong constraints on early Galactic nucleosynthesis 
and CCSNe models.
Previous Galactic chemical evolution  models employing various proportions 
of CCSNe have had difficulty 
in achieving the observed levels of Ti.  
Our work might suggest more energetic CCSNe, 
i.e., hypernovae, could have been 
more common, and thus responsible for this synthesis early in the history 
of the Galaxy.

Finally, our detailed neutral and ionized Co abundance measurements indicate 
that there are overabundances only when studying Co I lines. 
We suggest that [Co/Fe] values given by the ionized species are correct 
and there are no overabundances of this iron-peak element in very 
low-metallicity stars.

\acknowledgments

We thank our colleagues for help and advice, including X.\ Ou.
\textbf{We are also grateful to the referee for providing useful comments 
that helped improve this paper.}
This work has been supported in part by NASA grant NNX16AE96G (J.E.L.),
by National Science Foundation (NSF) 
grants AST-1516182 and AST-1814512 (J.E.L. and E.D.H.),  AST-1616040 (C.S.)
and  AST-1613536 and
AST-1815403 (IUR). 
JJC and IUR were supported in part by the
JINA Center for the Evolution of the Elements,
supported by the NSF under Grant No. PHY-1430152.

\facility{HST (STIS), Keck I (HIRES), McDonald 2.7m Smith (Tull), VLT (UVES)}

\software{\textbf{linemake (https://github.com/vmplacco/linemake), MOOG 
(\citealt{sneden73,sobeck11})}}

\clearpage
\bibliographystyle{apj}
\bibliography{totbib}

\begin{thebibliography}{}
\expandafter\ifx\csname natexlab\endcsname\relax\def\natexlab#1{#1}\fi

\bibitem[{{Andrievsky} {et~al.}(2018){Andrievsky}, {Bonifacio}, {Caffau},
  {Korotin}, {Spite}, {Spite}, {Sbordone}, \& {Zhukova}}]{andrievsky18b}
{Andrievsky}, S., {Bonifacio}, P., {Caffau}, E., {et~al.} 2018, \mnras, 473,
  3377

\bibitem[{{Asplund} {et~al.}(2009){Asplund}, {Grevesse}, {Sauval}, \&
  {Scott}}]{asplund09}
{Asplund}, M., {Grevesse}, N., {Sauval}, A.~J., \& {Scott}, P. 2009, ARA\&A,
  47, 481

\bibitem[{{Barklem} {et~al.}(2005){Barklem}, {Christlieb}, {Beers}, {Hill},
  {Bessell}, {Holmberg}, {Marsteller}, {Rossi}, {Zickgraf}, \&
  {Reimers}}]{barklem05}
{Barklem}, P.~S., {Christlieb}, N., {Beers}, T.~C., {et~al.} 2005, A\&A, 439,
  129

\bibitem[{{Bergemann} \& {Gehren}(2008)}]{bergemann08}
{Bergemann}, M., \& {Gehren}, T. 2008, A\&A, 492, 823

\bibitem[{{Bergemann} {et~al.}(2010){Bergemann}, {Pickering}, \&
  {Gehren}}]{bergemann10}
{Bergemann}, M., {Pickering}, J.~C., \& {Gehren}, T. 2010, \mnras, 401, 1334

\bibitem[{{Boesgaard} {et~al.}(2011){Boesgaard}, {Rich}, {Levesque}, \&
  {Bowler}}]{boesgaard11}
{Boesgaard}, A.~M., {Rich}, J., {Levesque}, E.~M., \& {Bowler}, B.~P. 2011,
  ApJ, 743, 140

\bibitem[{{Bonifacio} {et~al.}(2007){Bonifacio}, {Molaro}, {Sivarani},
  {Cayrel}, {Spite}, {Spite}, {Plez}, {Andersen}, {Barbuy}, {Beers}, {Depagne},
  {Hill}, {Fran{\c{c}}ois}, {Nordstr{\"o}m}, \& {Primas}}]{bonifacio07}
{Bonifacio}, P., {Molaro}, P., {Sivarani}, T., {et~al.} 2007, \aap, 462, 851

\bibitem[{{Cayrel} {et~al.}(2004){Cayrel}, {Depagne}, {Spite}, {Hill}, {Spite},
  {Fran{\c c}ois}, {Plez}, {Beers}, {Primas}, {Andersen}, {Barbuy},
  {Bonifacio}, {Molaro}, \& {Nordstr{\"o}m}}]{cayrel04}
{Cayrel}, R., {Depagne}, E., {Spite}, M., {et~al.} 2004, A\&A, 416, 1117

\bibitem[{{Chieffi} \& {Limongi}(2013)}]{chieffi13}
{Chieffi}, A., \& {Limongi}, M. 2013, ApJ, 764, 21

\bibitem[{{Cohen} {et~al.}(2008){Cohen}, {Christlieb}, {McWilliam}, {Shectman},
  {Thompson}, {Melendez}, {Wisotzki}, \& {Reimers}}]{cohen08}
{Cohen}, J.~G., {Christlieb}, N., {McWilliam}, A., {et~al.} 2008, \apj, 672,
  320

\bibitem[{{Cohen} {et~al.}(2004){Cohen}, {Christlieb}, {McWilliam}, {Shectman},
  {Thompson}, {Wasserburg}, {Ivans}, {Dehn}, {Karlsson}, \&
  {Melendez}}]{cohen04}
---. 2004, ApJ, 612, 1107

\bibitem[{{Cowan} {et~al.}(2019){Cowan}, {Sneden}, {Lawler}, {Aprahamian},
  {Wiescher}, {Langanke}, {Mart{\'\i}nez-Pinedo}, \& {Thielemann}}]{cowan19}
{Cowan}, J.~J., {Sneden}, C., {Lawler}, J.~E., {et~al.} 2019, arXiv e-prints,
  arXiv:1901.01410

\bibitem[{{Curtis} {et~al.}(2019){Curtis}, {Ebinger}, {Frohlich}, {Hempel},
  {Perego}, \& {Liebendorfer}}]{curtis19}
{Curtis}, S., {Ebinger}, K., {Frohlich}, C., {et~al.} 2019, ApJ, 870, 2

\bibitem[{{Dekker} {et~al.}(2000){Dekker}, {D'Odorico}, {Kaufer}, {Delabre}, \&
  {Kotzlowski}}]{dekker00}
{Dekker}, H., {D'Odorico}, S., {Kaufer}, A., {Delabre}, B., \& {Kotzlowski}, H.
  2000, in Society of Photo-Optical Instrumentation Engineers (SPIE) Conference
  Series, Vol. 4008, \procspie, ed. M.~{Iye} \& A.~F. {Moorwood}, 534--545

\bibitem[{{Den Hartog} {et~al.}(2019){Den Hartog}, {Lawler}, {Sneden}, {Cowan},
  \& {Brukhovesky}}]{denhartog19}
{Den Hartog}, E.~A., {Lawler}, J.~E., {Sneden}, C., {Cowan}, J.~J., \&
  {Brukhovesky}, A. 2019, ApJS, 243, 33

\bibitem[{{Den Hartog} {et~al.}(2011){Den Hartog}, {Lawler}, {Sobeck},
  {Sneden}, \& {Cowan}}]{denhartog11}
{Den Hartog}, E.~A., {Lawler}, J.~E., {Sobeck}, J.~S., {Sneden}, C., \&
  {Cowan}, J.~J. 2011, ApJS, 194, 35

\bibitem[{{Ezzeddine} {et~al.}(2019){Ezzeddine}, {Frebel}, {Roederer},
  {Tominaga}, {Tumlinson}, {Ishigaki}, {Nomoto}, {Placco}, \&
  {Aoki}}]{ezzeddine19}
{Ezzeddine}, R., {Frebel}, A., {Roederer}, I.~U., {et~al.} 2019, \apj, 876, 97

\bibitem[{{Fedchak} \& {Lawler}(1999)}]{fedchak99}
{Fedchak}, J.~A., \& {Lawler}, J.~E. 1999, ApJ, 523, 734

\bibitem[{{Grevesse} {et~al.}(2015){Grevesse}, {Scott}, {Asplund}, \&
  {Sauval}}]{grevesse15}
{Grevesse}, N., {Scott}, P., {Asplund}, M., \& {Sauval}, A.~J. 2015, A\&A, 573,
  A27

\bibitem[{{Heger} \& {Woosley}(2010)}]{heger10}
{Heger}, A., \& {Woosley}, S.~E. 2010, ApJ, 724, 341

\bibitem[{{Hosford} {et~al.}(2009){Hosford}, {Ryan}, {Garc{\'{\i}}a P{\'e}rez},
  {Norris}, \& {Olive}}]{hosford09}
{Hosford}, A., {Ryan}, S.~G., {Garc{\'{\i}}a P{\'e}rez}, A.~E., {Norris},
  J.~E., \& {Olive}, K.~A. 2009, A\&A, 493, 601

\bibitem[{{Kimble} {et~al.}(1998){Kimble}, {Woodgate}, {Bowers}, {Kraemer},
  {Kaiser}, {Gull}, {Heap}, {Danks}, {Boggess}, {Green}, {Hutchings},
  {Jenkins}, {Joseph}, {Linsky}, {Maran}, {Moos}, {Roesler}, {Timothy},
  {Weistrop}, {Grady}, {Loiacono}, {Brown}, {Brumfield}, {Content}, {Feinberg},
  {Isaacs}, {Krebs}, {Krueger}, {Melcher}, {Rebar}, {Vitagliano}, {Yagelowich},
  {Meyer}, {Hood}, {Argabright}, {Becker}, {Bottema}, {Breyer}, {Bybee},
  {Christon}, {Delamere}, {Dorn}, {Downey}, {Driggers}, {Ebbets}, {Gallegos},
  {Garner}, {Hetlinger}, {Lettieri}, {Ludtke}, {Michika}, {Nyquist}, {Rose},
  {Stocker}, {Sullivan}, {Van Houten}, {Woodruff}, {Baum}, {Hartig}, {Balzano},
  {Biagetti}, {Blades}, {Bohlin}, {Clampin}, {Doxsey}, {Ferguson},
  {Goudfrooij}, {Hulbert}, {Kutina}, {McGrath}, {Lindler}, {Beck}, {Feggans},
  {Plait}, {Sandoval}, {Hill}, {Collins}, {Cornett}, {Fowler}, {Hill},
  {Landsman}, {Malumuth}, {Standley}, {Blouke}, {Grusczak}, {Reed}, {Robinson},
  {Valenti}, \& {Wolfe}}]{kimble98}
{Kimble}, R.~A., {Woodgate}, B.~E., {Bowers}, C.~W., {et~al.} 1998, \apjl, 492,
  L83

\bibitem[{{Kobayashi} {et~al.}(2011){Kobayashi}, {Karakas}, \&
  {Umeda}}]{kobayashi11b}
{Kobayashi}, C., {Karakas}, A.~I., \& {Umeda}, H. 2011, MNRAS, 414, 3231

\bibitem[{{Kobayashi} {et~al.}(2006){Kobayashi}, {Umeda}, {Nomoto}, {Tominaga},
  \& {Ohkubo}}]{kobayashi06}
{Kobayashi}, C., {Umeda}, H., {Nomoto}, K., {Tominaga}, N., \& {Ohkubo}, T.
  2006, ApJ, 653, 1145

\bibitem[{Kramida {et~al.}(2018)Kramida, {Yu.~Ralchenko}, Reader, \& {and NIST
  ASD Team}}]{kramida18}
Kramida, A., {Yu.~Ralchenko}, Reader, J., \& {and NIST ASD Team}. 2018, {NIST
  Atomic Spectra Database (ver. 5.6.1), [Online]. Available:
  {\tt{https://physics.nist.gov/asd}} [2016, January 31]. National Institute of
  Standards and Technology, Gaithersburg, MD.}

\bibitem[{{Kurucz}(2011)}]{kurucz11}
{Kurucz}, R.~L. 2011, Canadian Journal of Physics, 89, 417

\bibitem[{{Lai} {et~al.}(2008){Lai}, {Bolte}, {Johnson}, {Lucatello}, {Heger},
  \& {Woosley}}]{lai08}
{Lai}, D.~K., {Bolte}, M., {Johnson}, J.~A., {et~al.} 2008, ApJ, 681, 1524

\bibitem[{{Lawler} {et~al.}(2018){Lawler}, {Feigenson}, {Sneden}, {Cowan}, \&
  {Nave}}]{lawler18}
{Lawler}, J.~E., {Feigenson}, T., {Sneden}, C., {Cowan}, J.~J., \& {Nave}, G.
  2018, \apjs, 238, 7

\bibitem[{{Lawler} {et~al.}(2013){Lawler}, {Guzman}, {Wood}, {Sneden}, \&
  {Cowan}}]{lawler13}
{Lawler}, J.~E., {Guzman}, A., {Wood}, M.~P., {Sneden}, C., \& {Cowan}, J.~J.
  2013, ApJS, 205, 11

\bibitem[{{Lawler} {et~al.}(2019){Lawler}, {Hala}, {Sneden}, {Nave}, {Wood}, \&
  {Cowan}}]{lawler19}
{Lawler}, J.~E., {Hala}, {Sneden}, C., {et~al.} 2019, \apjs, 241, 21

\bibitem[{{Lawler} {et~al.}(2015){Lawler}, {Sneden}, \& {Cowan}}]{lawler15}
{Lawler}, J.~E., {Sneden}, C., \& {Cowan}, J.~J. 2015, ApJS, 220, 13

\bibitem[{{Lawler} {et~al.}(2009){Lawler}, {Sneden}, {Cowan}, {Ivans}, \& {Den
  Hartog}}]{lawler09}
{Lawler}, J.~E., {Sneden}, C., {Cowan}, J.~J., {Ivans}, I.~I., \& {Den Hartog},
  E.~A. 2009, ApJS, 182, 51

\bibitem[{{Lawler} {et~al.}(2017){Lawler}, {Sneden}, {Nave}, {Den Hartog},
  {Emraho{\u g}lu}, \& {Cowan}}]{lawler17}
{Lawler}, J.~E., {Sneden}, C., {Nave}, G., {et~al.} 2017, ApJS, 228, 10

\bibitem[{{Lawler} {et~al.}(2014){Lawler}, {Wood}, {Den Hartog}, {Feigenson},
  {Sneden}, \& {Cowan}}]{lawler14}
{Lawler}, J.~E., {Wood}, M.~P., {Den Hartog}, E.~A., {et~al.} 2014, ApJS, 215,
  20

\bibitem[{{Limongi} {et~al.}(2000){Limongi}, {Straniero}, \&
  {Chieffi}}]{limongi00}
{Limongi}, M., {Straniero}, O., \& {Chieffi}, A. 2000, ApJS, 129, 625

\bibitem[{{McWilliam} {et~al.}(1995){McWilliam}, {Preston}, {Sneden}, \&
  {Searle}}]{mcwilliam95}
{McWilliam}, A., {Preston}, G.~W., {Sneden}, C., \& {Searle}, L. 1995, AJ, 109,
  2757

\bibitem[{{Mel{\'e}ndez} \& {Barbuy}(2009)}]{melendez09}
{Mel{\'e}ndez}, J., \& {Barbuy}, B. 2009, A\&A, 497, 611

\bibitem[{{Nomoto} {et~al.}(2013){Nomoto}, {Kobayashi}, \&
  {Tominaga}}]{nomoto13}
{Nomoto}, K., {Kobayashi}, C., \& {Tominaga}, N. 2013, \araa, 51, 457

\bibitem[{{Nordlander} {et~al.}(2019){Nordlander}, {Bessell}, {Da Costa},
  {Mackey}, {Asplund}, {Casey}, {Chiti}, {Ezzeddine}, {Frebel}, {Lind},
  {Marino}, {Murphy}, {Norris}, {Schmidt}, \& {Yong}}]{nordlander19}
{Nordlander}, T., {Bessell}, M.~S., {Da Costa}, G.~S., {et~al.} 2019, \mnras,
  488, L109

\bibitem[{{Pignatari} {et~al.}(2016){Pignatari}, {Herwig}, {Hirschi},
  {Bennett}, {Rockefeller}, {Fryer}, {Timmes}, {Ritter}, {Heger}, {Jones},
  {Battino}, {Dotter}, {Trappitsch}, {Diehl}, {Frischknecht}, {Hungerford},
  {Magkotsios}, {Travaglio}, \& {Young}}]{pignatari16}
{Pignatari}, M., {Herwig}, F., {Hirschi}, R., {et~al.} 2016, \apjs, 225, 24

\bibitem[{{Rauscher} {et~al.}(2002){Rauscher}, {Heger}, {Hoffman}, \&
  {Woosley}}]{rauscher02}
{Rauscher}, T., {Heger}, A., {Hoffman}, R.~D., \& {Woosley}, S.~E. 2002, ApJ,
  576, 323

\bibitem[{{Roederer} \& {Barklem}(2018)}]{roederer18a}
{Roederer}, I.~U., \& {Barklem}, P.~S. 2018, \apj, 857, 2

\bibitem[{{Roederer} {et~al.}(2014{\natexlab{a}}){Roederer}, {Preston},
  {Thompson}, {Shectman}, {Sneden}, {Burley}, \& {Kelson}}]{roederer14}
{Roederer}, I.~U., {Preston}, G.~W., {Thompson}, I.~B., {et~al.}
  2014{\natexlab{a}}, AJ, 147, 136

\bibitem[{{Roederer} {et~al.}(2018){Roederer}, {Sneden}, {Lawler}, {Sobeck},
  {Cowan}, \& {Boesgaard}}]{roederer18}
{Roederer}, I.~U., {Sneden}, C., {Lawler}, J.~E., {et~al.} 2018, ApJ, 860, 125

\bibitem[{{Roederer} {et~al.}(2014{\natexlab{b}}){Roederer}, {Schatz},
  {Lawler}, {Beers}, {Cowan}, {Frebel}, {Ivans}, {Sneden}, \&
  {Sobeck}}]{roederer14b}
{Roederer}, I.~U., {Schatz}, H., {Lawler}, J.~E., {et~al.} 2014{\natexlab{b}},
  \apj, 791, 32

\bibitem[{{Ryan} {et~al.}(1996){Ryan}, {Norris}, \& {Beers}}]{ryan96}
{Ryan}, S.~G., {Norris}, J.~E., \& {Beers}, T.~C. 1996, ApJ, 471, 254

\bibitem[{{Ryan} {et~al.}(1999){Ryan}, {Norris}, \& {Beers}}]{ryan99}
---. 1999, \apj, 523, 654

\bibitem[{{Ryan} {et~al.}(1991){Ryan}, {Norris}, \& {Bessell}}]{ryan91}
{Ryan}, S.~G., {Norris}, J.~E., \& {Bessell}, M.~S. 1991, \aj, 102, 303

\bibitem[{{Shi} {et~al.}(2018){Shi}, {Yan}, {Zhou}, \& {Zhao}}]{shi18}
{Shi}, J.~R., {Yan}, H.~L., {Zhou}, Z.~M., \& {Zhao}, G. 2018, \apj, 862, 71

\bibitem[{{Sneden}(1973)}]{sneden73}
{Sneden}, C. 1973, ApJ, 184, 839

\bibitem[{{Sneden} {et~al.}(2016){Sneden}, {Cowan}, {Kobayashi}, {Pignatari},
  {Lawler}, {Den Hartog}, \& {Wood}}]{sneden16}
{Sneden}, C., {Cowan}, J.~J., {Kobayashi}, C., {et~al.} 2016, ApJ, 817, 53

\bibitem[{{Sneden} {et~al.}(2009){Sneden}, {Lawler}, {Cowan}, {Ivans}, \& {Den
  Hartog}}]{sneden09}
{Sneden}, C., {Lawler}, J.~E., {Cowan}, J.~J., {Ivans}, I.~I., \& {Den Hartog},
  E.~A. 2009, ApJS, 182, 80

\bibitem[{{Sobeck} {et~al.}(2007){Sobeck}, {Lawler}, \& {Sneden}}]{sobeck07}
{Sobeck}, J.~S., {Lawler}, J.~E., \& {Sneden}, C. 2007, ApJ, 667, 1267

\bibitem[{{Sobeck} {et~al.}(2011){Sobeck}, {Kraft}, {Sneden}, {Preston},
  {Cowan}, {Smith}, {Thompson}, {Shectman}, \& {Burley}}]{sobeck11}
{Sobeck}, J.~S., {Kraft}, R.~P., {Sneden}, C., {et~al.} 2011, AJ, 141, 175

\bibitem[{{Tull} {et~al.}(1995){Tull}, {MacQueen}, {Sneden}, \&
  {Lambert}}]{tull95}
{Tull}, R.~G., {MacQueen}, P.~J., {Sneden}, C., \& {Lambert}, D.~L. 1995,
  \pasp, 107, 251

\bibitem[{{Umeda} \& {Nomoto}(2002)}]{umeda02}
{Umeda}, H., \& {Nomoto}, K. 2002, ApJ, 565, 385

\bibitem[{{Vogt} {et~al.}(1994){Vogt}, {Allen}, {Bigelow}, {Bresee}, {Brown},
  {Cantrall}, {Conrad}, {Couture}, {Delaney}, {Epps}, {Hilyard}, {Hilyard},
  {Horn}, {Jern}, {Kanto}, {Keane}, {Kibrick}, {Lewis}, {Osborne},
  {Pardeilhan}, {Pfister}, {Ricketts}, {Robinson}, {Stover}, {Tucker}, {Ward},
  \& {Wei}}]{vogt94}
{Vogt}, S.~S., {Allen}, S.~L., {Bigelow}, B.~C., {et~al.} 1994, in Society of
  Photo-Optical Instrumentation Engineers (SPIE) Conference Series, Vol. 2198,
  Instrumentation in Astronomy VIII, ed. D.~L. {Crawford} \& E.~R. {Craine},
  362

\bibitem[{{Wallerstein} \& {Helfer}(1959)}]{wallerstein59}
{Wallerstein}, G., \& {Helfer}, H.~L. 1959, ApJ, 129, 720

\bibitem[{{Wood} {et~al.}(2014a){Wood}, {Lawler}, {Den Hartog}, {Sneden}, \&
  {Cowan}}]{wood14a}
{Wood}, M.~P., {Lawler}, J.~E., {Den Hartog}, E.~A., {Sneden}, C., \& {Cowan},
  J.~J. 2014a, ApJS, 214, 18

\bibitem[{{Wood} {et~al.}(2013){Wood}, {Lawler}, {Sneden}, \& {Cowan}}]{wood13}
{Wood}, M.~P., {Lawler}, J.~E., {Sneden}, C., \& {Cowan}, J.~J. 2013, ApJS,
  208, 27

\bibitem[{{Wood} {et~al.}(2014b){Wood}, {Lawler}, {Sneden}, \&
  {Cowan}}]{wood14b}
---. 2014b, ApJS, 211, 20

\bibitem[{{Wood} {et~al.}(2018){Wood}, {Sneden}, {Lawler}, {Den Hartog},
  {Cowan}, \& {Nave}}]{wood18}
{Wood}, M.~P., {Sneden}, C., {Lawler}, J.~E., {et~al.} 2018, ApJS, 234, 25

\bibitem[{{Woodgate} {et~al.}(1998){Woodgate}, {Kimble}, {Bowers}, {Kraemer},
  {Kaiser}, {Danks}, {Grady}, {Loiacono}, {Brumfield}, {Feinberg}, {Gull},
  {Heap}, {Maran}, {Lindler}, {Hood}, {Meyer}, {Vanhouten}, {Argabright},
  {Franka}, {Bybee}, {Dorn}, {Bottema}, {Woodruff}, {Michika}, {Sullivan},
  {Hetlinger}, {Ludtke}, {Stocker}, {Delamere}, {Rose}, {Becker}, {Garner},
  {Timothy}, {Blouke}, {Joseph}, {Hartig}, {Green}, {Jenkins}, {Linsky},
  {Hutchings}, {Moos}, {Boggess}, {Roesler}, \& {Weistrop}}]{woodgate98}
{Woodgate}, B.~E., {Kimble}, R.~A., {Bowers}, C.~W., {et~al.} 1998, \pasp, 110,
  1183

\bibitem[{{Woosley} \& {Weaver}(1995)}]{woosley95}
{Woosley}, S.~E., \& {Weaver}, T.~A. 1995, ApJS, 101, 181

\bibitem[{{Yong} {et~al.}(2013){Yong}, {Norris}, {Bessell}, {Christlieb},
  {Asplund}, {Beers}, {Barklem}, {Frebel}, \& {Ryan}}]{yong13}
{Yong}, D., {Norris}, J.~E., {Bessell}, M.~S., {et~al.} 2013, ApJ, 762, 26

\bibitem[{{Yousefi} \& {Bernath}(2018)}]{yousefi18}
{Yousefi}, M., \& {Bernath}, P.~F. 2018, \apjs, 237, 8

\end{thebibliography}


\clearpage

\clearpage

\clearpage

\clearpage

\clearpage                                                    

\clearpage

\clearpage

\clearpage

\clearpage

\clearpage

\clearpage


\clearpage
\begin{center}
\begin{deluxetable}{lcccc}
\tabletypesize{\footnotesize}
\tablewidth{0pt}
\tablecaption{Model Atmospheric Parameters\label{tab-models}}
\tablecolumns{5}
\tablehead{
\colhead{Star}                &
\colhead{\teff}               &
\colhead{\logg}               &
\colhead{\vmicro}             &
\colhead{[M/H]}               \\
\colhead{}                    &
\colhead{K}                   &
\colhead{}                    &
\colhead{\kmsec}              &
\colhead{}                     
}
\startdata
\bdzero\      &   6351   &   3.97   &   1.70   &   $-$2.90   \\
\bdone\       &   6405   &   4.04   &   1.60   &   $-$2.85   \\
\cdthree\     &   6625   &   4.29   &   1.60   &   $-$3.00   \\
\hdeight\     &   6300   &   4.00   &   1.50   &   $-$2.15   \\
\enddata                                                      
\end{deluxetable}                                             
\end{center}

\clearpage
\begin{center}
\begin{deluxetable}{rcrrrrrr}
\tabletypesize{\footnotesize}
\tablewidth{0pt}
\tablecaption{Line Parameters and Abundances\label{tab-lines}}
\tablecolumns{7}
\tablehead{
\colhead{$\lambda$}           &
\colhead{Species}             &
\colhead{$\chi$}              &
\colhead{log $gf$}            &
\colhead{log $\epsilon$}      &
\colhead{log $\epsilon$}      &
\colhead{log $\epsilon$}      \\
\colhead{\AA}                 &
\colhead{}                    &
\colhead{eV}                  &
\colhead{}                    &
\colhead{\bdzero}             &
\colhead{\bdone}              &
\colhead{\cdthree}             
}
\startdata
  2552.354 & \species{Sc}{ii}  &  0.022 &     0.05 &   0.50 &   0.65 &   0.45 \\
  2555.795 & \species{Sc}{ii}  &  0.000 &  $-$0.69 &   0.60 &   0.80 &   0.65 \\
  2563.190 & \species{Sc}{ii}  &  0.000 &  $-$0.57 &   0.40 &   0.75 &\nodata \\
  3353.724 & \species{Sc}{ii}  &  0.315 &     0.26 &   0.58 &   0.75 &   0.52 \\
  3359.678 & \species{Sc}{ii}  &  0.008 &  $-$0.75 &   0.70 &   0.75 &   0.65 \\
  3361.931 & \species{Sc}{ii}  &  0.000 &  $-$0.72 &   0.65 &   0.85 &   0.65 \\
  3368.936 & \species{Sc}{ii}  &  0.008 &  $-$0.39 &   0.60 &   0.77 &   0.58 \\
  3535.714 & \species{Sc}{ii}  &  0.315 &  $-$0.46 &   0.55 &   0.75 &   0.50 \\
  3567.696 & \species{Sc}{ii}  &  0.000 &  $-$0.47 &   0.58 &   0.74 &   0.46 \\
  3572.526 & \species{Sc}{ii}  &  0.022 &     0.27 &   0.61 &   0.81 &   0.57 \\
  3576.340 & \species{Sc}{ii}  &  0.008 &     0.01 &   0.55 &   0.74 &   0.52 \\
  3580.925 & \species{Sc}{ii}  &  0.000 &  $-$0.14 &   0.55 &   0.74 &   0.48 \\
  3589.632 & \species{Sc}{ii}  &  0.008 &  $-$0.57 &   0.60 &   0.77 &   0.53 \\
\enddata
\tablecomments{The complete version of this table is available
in the online edition of the journal.
An abbreviated version is shown here to illustrate its form and content.}
\end{deluxetable}
\end{center}

\clearpage
\begin{center}
\begin{deluxetable}{lcc@{\extracolsep{0.1in}}cccccccccccc}
\tabletypesize{\footnotesize}
\tablewidth{0pt}
\tablecaption{Mean Abundances\label{tab-abmeans}}
\tablecolumns{15}
\tablehead{
\multicolumn{1}{c}{}                      &
\multicolumn{2}{c}{Sun}                   &
\multicolumn{4}{c}{\bdzero}               &
\multicolumn{4}{c}{\bdone}                &
\multicolumn{4}{c}{\cdthree}              \\
\cline{2-3} \cline{4-7} \cline{8-11} \cline{12-15}
\multicolumn{1}{c}{El}                    &
\multicolumn{1}{c}{log $\epsilon$}        &
\multicolumn{1}{c}{source}                &
\multicolumn{1}{c}{log $\epsilon$}        &
\multicolumn{1}{c}{$\sigma$}              &
\multicolumn{1}{c}{num}                   &
\multicolumn{1}{c}{[X/H]}                 &
\multicolumn{1}{c}{log $\epsilon$}        &
\multicolumn{1}{c}{$\sigma$}              &
\multicolumn{1}{c}{num}                   &
\multicolumn{1}{c}{[X/H]}                 &
\multicolumn{1}{c}{log $\epsilon$}        &
\multicolumn{1}{c}{$\sigma$}              &
\multicolumn{1}{c}{num}                   &
\multicolumn{1}{c}{[X/H]}
}
\startdata
\multicolumn{15}{c}{Neutral Species} \\
Sc    &   3.15  &      1  & \nodata  & \nodata  & \nodata  & \nodata  & \nodata  & \nodata  & \nodata  & \nodata  & \nodata  & \nodata  & \nodata  & \nodata \\ Ti    &   4.97  &      2  &    2.71  &    0.05  &      24  & $-$2.26  &    2.79  &    0.06  &      17  & $-$2.18  &    2.71  &    0.08  &      10  & $-$2.26 \\
V     &   3.96  &      4  & \nodata  & \nodata  & \nodata  & \nodata  & \nodata  & \nodata  & \nodata  & \nodata  & \nodata  & \nodata  & \nodata  & \nodata \\ Cr    &   5.64  &      6  &    2.74  &    0.09  &      13  & $-$2.90  &    2.81  &    0.13  &      12  & $-$2.83  &    2.69  &    0.10  &      10  & $-$2.95 \\ Mn    &   5.45  &      8  &    2.23  &    0.07  &       5  & $-$3.22  &    2.21  &    0.08  &       5  & $-$3.24  &    2.12  &    0.06  &       3  & $-$3.33 \\ Fe    &   7.50  &     10  &    4.61  &    0.07  &     230  & $-$2.89  &    4.66  &    0.09  &     243  & $-$2.84  &    4.52  &    0.14  &     194  & $-$2.98 \\ Co    &   4.96  &     11  &    2.41  &    0.11  &      36  & $-$2.55  &    2.43  &    0.11  &      31  & $-$2.53  &    2.45  &    0.12  &      28  & $-$2.51 \\
Ni    &   6.28  &     13  &    3.39  &    0.06  &      45  & $-$2.89  &    3.43  &    0.09  &      43  & $-$2.85  &    3.39  &    0.08  &      35  & $-$2.89 \\
Cu    &   4.18  &     15  &    0.71  &   (0.1)\tablenotemark{a} &       2  & $-$3.47  &    0.71  &   (0.1)\tablenotemark{a}  &       2  & $-$3.47  &    0.51  &    0.14  &       2  & $-$3.67 \\
Zn    &   4.56  &     15  &    1.90  &   (0.1)\tablenotemark{a}  &       2  & $-$2.66  &    2.10  &    0.14  &       2  & $-$2.46  & \nodata  & \nodata  & \nodata  & \nodata \\
\multicolumn{15}{c}{Ionized Species} \\
Sc    &   3.16  &      1  &    0.57  &    0.07  &      27  & $-$2.59  &    0.77  &    0.06  &      27  & $-$2.39  &    0.53  &    0.07  &      19  & $-$2.63 \\ Ti    &   4.98  &      3  &    2.67  &    0.10  &     108  & $-$2.31  &    2.71  &    0.11  &      99  & $-$2.27  &    2.53  &    0.11  &      72  & $-$2.45 \\
V     &   3.95  &      5  &    1.51  &    0.10  &      47  & $-$2.44  &    1.62  &    0.11  &      41  & $-$2.33  &    1.41  &    0.09  &      30  & $-$2.54 \\
Cr    &   5.62  &      7  &    2.92  &    0.06  &      50  & $-$2.70  &    2.97  &    0.06  &      57  & $-$2.65  &    2.79  &    0.07  &      56  & $-$2.83 \\
Mn    &   5.45  &      9  &    2.41  &    0.03  &      10  & $-$3.04  &    2.43  &    0.05  &       6  & $-$3.02  &    2.28  &    0.07  &       6  & $-$3.17 \\
Fe    &   7.50  &     10  &    4.72  &    0.10  &      68  & $-$2.78  &    4.77  &    0.09  &      70  & $-$2.73  &    4.60  &    0.08  &      61  & $-$2.90 \\
Co    &   4.96  &     12  &    2.04  &    0.08  &      21  & $-$2.92  &    2.08  &    0.10  &      19  & $-$2.88  &    1.96  &    0.11  &      16  & $-$3.00 \\ Ni    &   6.28  &     14  &    3.39  &    0.19  &       8  & $-$2.89  &    3.39  &    0.15  &       8  & $-$2.89  &    3.31  &    0.13  &       7  & $-$2.97 \\
Cu    &   4.18  &     16  & \nodata  & \nodata  & \nodata  & \nodata  & \nodata  & \nodata  & \nodata  & \nodata  & \nodata  & \nodata  & \nodata  & \nodata \\ Zn    &   4.61  &     16  & \nodata  & \nodata  & \nodata  & \nodata  & \nodata  & \nodata  & \nodata  & \nodata  & \nodata  & \nodata  & \nodata  & \nodata \\
\enddata
\tablenotetext{a}{The formal $\sigma$ value is unrealistically close to zero,
                  so the table value is arbitrarily set to 0.1}
\tablerefs{The citations to solar 
abundance sources enumerated in the third column are:
            [1] \cite{lawler19}; [2] \cite{lawler13}; [3] \cite{wood13};
            [4] \cite{lawler14}; [5] \cite{wood14a}, [6] \cite{sobeck07};
            [7] \cite{lawler17}; [8] \cite{sneden16}; [9] assumed from 
 the \species{Mn}{i}
            solar abundance \citep{sneden16};
            [10] \cite{asplund09}; [11] \cite{lawler15}; [12] assumed 
 from the
            \species{Co}{i} solar abundance \citep{lawler15};
            [13] \cite{wood14b}; [14] assumed from the \species{Ni}{i} 
 abundance;
            [15] \citep{grevesse15}; [16] assumed from the 
\species{Cu}{i} and
            \species{Zn}{i} solar abundances \citep{grevesse15}.}
\end{deluxetable}
\end{center}

\clearpage
\begin{center}
\begin{deluxetable}{lrrrrrrrrr}
\tabletypesize{\footnotesize}
\tablewidth{0pt}
\tablecaption{Final Abundance Ratios\label{tab-abratios}}
\tablecolumns{10}
\tablehead{
\colhead{El}                                           &
\colhead{[\species{X}{i}/Fe]}                          &
\colhead{[\species{X}{i}/Fe]}                          &
\colhead{[\species{X}{i}/Fe]}                          &
\colhead{$\langle$[\species{X}{i}/Fe]$\rangle$}        &
\colhead{[\species{X}{ii}/Fe]}                         &
\colhead{[\species{X}{ii}/Fe]}                         &
\colhead{[\species{X}{ii}/Fe]}                         &
\colhead{$\langle$[\species{X}{ii}/Fe]$\rangle$}       &
\colhead{$\langle$[X/Fe]$\rangle$}                     \\
\colhead{}                                             &
\colhead{\bdzero}                                      &
\colhead{\bdone}                                       &
\colhead{\cdthree}                                     &
\colhead{}                                             &
\colhead{\bdzero}                                      &
\colhead{\bdone}                                       &
\colhead{\cdthree}                                     &
\colhead{}                                             &
\colhead{El}                                           
}
\startdata
       Sc   &   \nodata   &   \nodata   &   \nodata   &  \nodata    &       0.19  &       0.34  &       0.27  &       0.27  &       0.27 \\   
       Ti   &       0.63  &       0.66  &       0.72  &       0.67  &       0.47  &       0.46  &       0.45  &       0.46  &       0.57 \\  
        V   &   \nodata   &   \nodata   &   \nodata   &  \nodata    &       0.34  &       0.40  &       0.36  &       0.37  &       0.37 \\ 
       Cr   &    $-$0.01  &       0.01  &       0.03  &       0.01  &       0.08  &       0.08  &       0.07  &       0.08  &            \\
   Cr-rev\tablenotemark{a} &       
         0.04  &       0.11  &       0.00  &       0.05  &   \nodata   &   \nodata   &   \nodata   &   \nodata   &       0.06 \\   
       Mn   &    $-$0.33  &    $-$0.40  &    $-$0.35  &    $-$0.36  &    $-$0.26  &    $-$0.29  &    $-$0.27  &    $-$0.27  &            \\
   Mn-rev\tablenotemark{b} &  
   $-$0.16  &    $-$0.22  &    $-$0.27  &    $-$0.22  &   \nodata   &   \nodata   &   \nodata   &   \nodata   &    $-$0.25 \\   
       Fe   &       0.00  &       0.00  &       0.00  &       0.00  &       0.00  &       0.00  &       0.00  &       0.00  &       0.00 \\
       Co   &       0.34  &       0.31  &       0.47  &       0.37  &    $-$0.14  &    $-$0.15  &    $-$0.10  &    $-$0.13  &    $-$0.13 \\         
       Ni   &       0.00  &    $-$0.01  &       0.09  &       0.03  &    $-$0.11  &    $-$0.16  &    $-$0.07  &    $-$0.11  &       0.03 \\
       Cu   &    $-$0.58  &    $-$0.63  &    $-$0.69  &    $-$0.63  &   \nodata   &   \nodata   &   \nodata   &   \nodata   &            \\
   Cu-rev\tablenotemark{c} &      
      $-$0.08  &    $-$0.13  &    $-$0.19  &    $-$0.13  &   \nodata   &   \nodata   &   \nodata   &   \nodata   &    $-$0.13 \\
       Zn   &       0.23  &       0.38  &   \nodata   &       0.31  &   \nodata   &   \nodata   &   \nodata   &   \nodata   &       0.31 \\
\enddata
\tablenotetext{a}{Cr-rev is the \species{Cr}{i} abundance neglecting the
                  transitions arising from the 0~eV ground state}
\tablenotetext{b}{Mn-rev is the \species{Mn}{i} abundance neglecting the
                  transitions arising from the 0~eV ground state}
\tablenotetext{c}{Cu-rev is the \species{Cu}{i} abundance with the addition
                  of 0.5~dex, roughly approximating an NLTE correction}
\end{deluxetable}
\end{center}

\end{document}